\documentclass[preprint]{jfm}

\usepackage[T1]{fontenc}

\lefttitle{M. de Wildt, A. Prosperetti, C. Diddens and D. Lohse}
\righttitle{Journal of Fluid Mechanics}

\title{Leidenfrost droplets: The roles of ambient humidity and internal droplet circulation}

\author{
    Maxim de Wildt\aff{1} 
    Andrea Prosperetti\aff{1,2,3}, 
    Christian Diddens\aff{1} 
    \and Detlef Lohse\aff{1,}\aff{4} 
    }

\affiliation{\aff{1}Physics of Fluids Department, Max-Planck Center Twente for Complex Fluid Dynamics and J.M. Burgers
Centre for Fluid Dynamics, University of Twente, PO Box 217, 7500AE Enschede, The Netherlands
\aff{2}Department of Mechanical and Aerospace Engineering, University of Houston, Houston, TX 77204, USA
\aff{3}Department of Mechanical Engineering, Johns Hopkins University, Baltimore, MD 21218, USA
\aff{4}Max-Planck Institute for Dynamics and Self-Organization, Am Faßberg 17, 37077 Göttingen, Germany
}

\corresau{
    Maxim de Wildt, \email{m.dewildt@utwente.nl}; 
    Christian Diddens, \email{c.diddens@utwente.nl}; 
    Detlef Lohse, \email{d.lohse@utwente.nl}
    }

\begin{document}
\maketitle

\begin{abstract}
A volatile droplet gently deposited on a superheated substrate can sit on a thin film of its own vapour, which prevents contact between the drop and surface. 
This phenomenon is called the Leidenfrost effect.
In this paper, through direct numerical simulations, we analyse characteristics of Leidenfrost water droplets with a single computational model over four decades of droplet radius, stitching together previous works in the limit of large and small droplets. 
Using the model, we show that the ambient humidity, an underappreciated factor in the Leidenfrost system, in combination with the flow in the drop has a significant impact on the geometry and drying kinetics.
Our results imply the inadequacies of commonly made assumptions of a pure vapour phase and an isothermal droplet. 
When modelling large Leidenfrost droplets with an axisymmetric model, large discrepancies between experiments and the computational results occur.
Through azimuthal stability analysis, we show that this is due to the unrealistic constraint of axisymmetry.
This finding is supported by 3D simulations of a simplified model.
Finally, some hypotheses are explored to account for the remaining discrepancies with experimental data.
\end{abstract}



\section{Introduction}
\label{sec:intro}

The Leidenfrost effect has been known for many centuries. It was first reported to be observed in the Netherlands by \citet{alma991015818599705181}.
It takes its name, however, from Johann Gottlob Leidenfrost, who in 1756 published an article reporting the fact that liquid droplets do not make contact with surfaces well above the boiling point of the liquid and have a lifetime much longer than expected \citep{leidenfrost1756aquae}. 
This phenomenon occurs because the droplet evaporates and sits on a thin layer of its own vapour, which creates sufficient pressure to balance the weight of the drop, thus preventing contact. 
Despite the discovery of the phenomenon nearly 300 years ago, as of today many research questions on this topic are still open.
For excellent recent reviews we refer to \citet{quere_leidenfrost_2013}, \citet{ajaev} and \citet{Stewart_2022}, which cover much of the work in the last two decades. 
There are many technological and industrial applications of the Leidenfrost effect which include metallurgy \citep{KARWA20131118}, spray cooling \citep{KIM2007753} and surface cleaning \citep{surf_cleaning}; most applications concern heat transfer in an attempt to limit the occurrence of the Leidenfrost effect \citep{Gu_recent_adv}.

The vapour layer is central to the rich portfolio of behaviours that liquids in the Leidenfrost state can exhibit, such as spontaneous beginning of trampolining/sustained bouncing \citep{graeber_leidenfrost_2021}, star-shaped oscillations \citep{brunet_star-drops_2011, ma_star-shaped_2017} or Leidenfrost wheels where droplets begin to roll \citep{bouillant_leidenfrost_2018}. 
Leidenfrost droplets can also levitate over hot liquids, with a wealth of research existing for droplets on liquid pools \citep{maquet_leidenfrost_2016,van_limbeek_asymptotic_2019}. 
There has also been much investigation into the dynamic Leidenfrost effect \citep{tran_drop_2012,shirota_dynamic_2016,chantelot_drop_2023}. However, this is outside the scope of this paper.

Accurate measurements of the geometry of Leidenfrost droplets began after the seminal paper from \citet{biance_leidenfrost_2003}, from which the first experimental measurements of the vapour layer thickness were made, with results ranging between 30-100$\upmu$m. 
They noted that droplets with initial radii $R\lesssim \ell_c$, where $\ell_c=\sqrt{\gamma/\rho_lg}$ is the capillary length, are quasi-spherical. 
In contrast, droplets larger than this formed puddles with height limited to $\approx2\ell_c$ and a characteristic concave vapour layer under the drop, uncovered by \citet{burton_geometry_2012} through laser interferometry. 
The upper bound of the size of these drops is limited by the formation of a `chimney' instability through the centre of the drop, which was investigated  theoretically by \citet{snoeijer_maximum_2009} and understood to be akin to a Rayleigh-Taylor instability occurring when the maximum radius of the disc formed by the drop reaches $\approx 4\ell_c$. 
In the case of very small droplets with $R\ll \ell_c$, the force exerted by the ejected vapour can be so strong as to lift them to significant heights above the plane before eventually disappearing \citep{celestini2012}. 
Other final fates may occur such as explosion when the drop liquid is not sufficiently clean \citep{lyu_final_2019,moreau_explosive_2019}. 

Lubrication models have been shown to very accurately model the shape of large Leidenfrost drops \citep{sobac_leidenfrost_2014, sobac_erratum_2021}, with the derived scaling laws for some features of the droplet geometry and evaporation agreeing well with experimental observations \citep{biance_leidenfrost_2003,POMEAU2012867, sobac_leidenfrost_2014}. 
However, these models neglect the flow inside the droplet and do not predict accurate lifetimes. The work of \citet{Chakraborty_Chubynsky_Sprittles_2022} took the lubrication models a step further by coupling shear from numerically solved Navier-Stokes equation in the droplet to the lubrication flow in the vapour layer. 
Although an improvement on the previous modelling, this model still assumes that the droplet effectively is present in a gaseous phase of pure vapour and, therefore, must be isothermal at thermodynamic equilibrium (cf. §\ref{ssec:jump}). 
In \citet{yim_leidenfrost_2022}, the authors applied an experimentally measured temperature gradient across the drop and investigated the onset of azimuthal symmetry breaking. However, the shear from the gas phase was not modelled. 
Instead, the dynamics were driven by thermal Marangoni flows. 
Indeed, from infrared measurements and particle image velocimetry (PIV) measurements from \citet{bouillant_leidenfrost_2018}, it is clear that Leidenfrost droplets experience significant cooling at their top ($\sim 10K$), which leads to thermal gradients along the droplet's surface and thus Marangoni flow, as well as strong internal flows on the order of $\mathrm{cm\, s^{-1}}$ in millimetric drops.

To help further understanding, direct numerical simulations encapsulating the phase change and interaction with a gas phase must be utilised. 
There have been many studies conducted in this fashion mainly concerning dynamic Leidenfrost scenarios, cf. \citet{rueda_villegas_direct_2017,wang_interface_2020,yuan_dynamical_2022,du_state_2024,zhao_direct_2025}, with many works making the simplifying assumptions of a pure vapour gaseous phase and isothermal droplet. 
There has been limited attention given to modelling quasi-stationary Leidenfrost droplets and clarifying the role of circulation in the droplet and vapour concentration outside the droplet across the entire droplet lifespan.

In this paper, we utilise a computational model to simulate Leidenfrost droplets across the entire span of their lifetime, avoiding some limitations of the aforementioned studies. 
Throughout this work, we explore the effects of ambient humidity and the role of circulation in the drop by using the viscosity of the drop as a tool to decouple the evaporation kinetics and internal motion in the droplet. 
Details of the mathematical model and numerical method are discussed in §\ref{sec:model} and §\ref{sec:method}, respectively. In §\ref{sec:shapes}, we go on to analyse the geometric characteristics of the Leidenfrost droplets and compare our results to the results from lubrication theory, as well as theoretical results from the Leidenfrost take-off limit $R\ll \ell_c$. 
In the same section, we obtain the very surprising and unintuitive result that models where internal drop circulation is included predict less accurate droplet shapes. 
The reason for this is hypothesised to be the onset of azimuthal instability of the system, which is determined in §\ref{sec:stab} and by 3D simulations shown to be the primary reason in the penultimate section of the paper.
Following this, in §\ref{sec:humcirc}, we provide insight into the impact which ambient humidity and droplet circulation have on various characteristics of the Leidenfrost drop. 
We then, in §\ref{sec:evap}, analyse the evaporation throughout the droplet lifetime.
Finally in §\ref{sec:extras}, we explore possible means to further improve the agreement of our results with experiments.
This includes thermal Marangoni effects and surfactant contamination as potential shortcomings of this model.

\section{Mathematical model}
\label{sec:model}

\subsection{Governing equations}
\label{ssec:gov}

\begin{figure}
  \centerline{\includegraphics[scale=0.55]{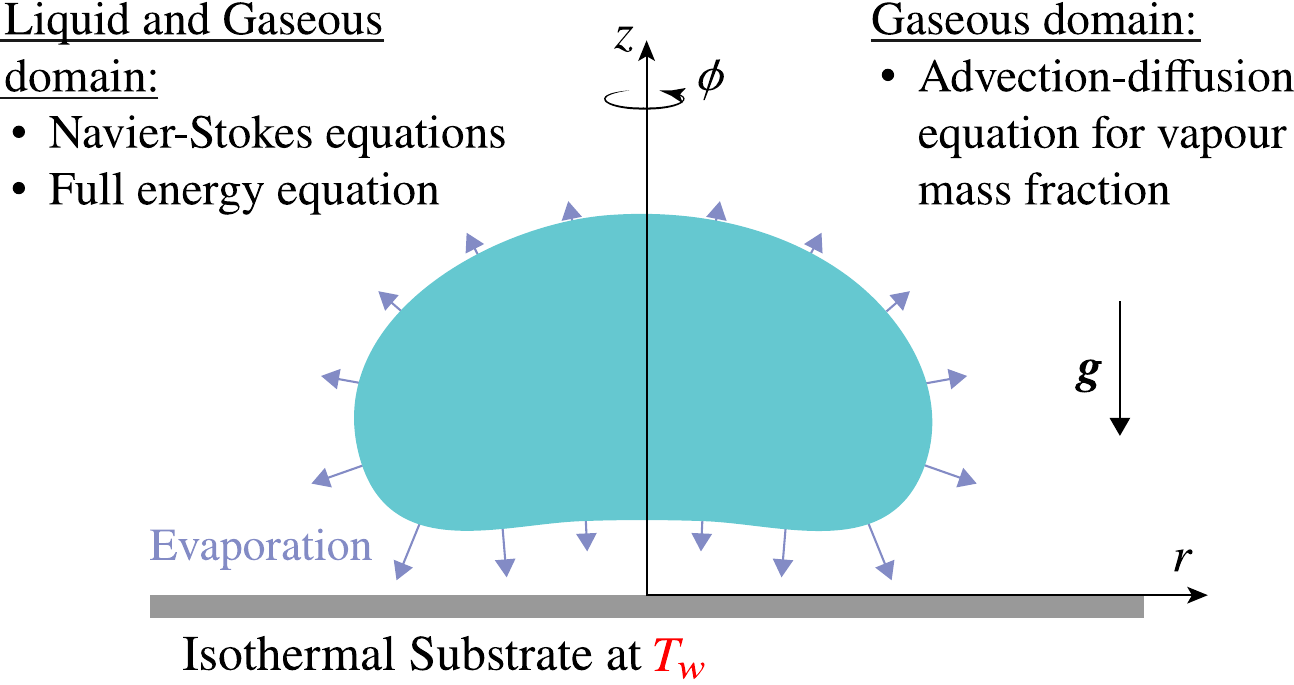}}
  \caption{Schematic of problem}
\label{fig:schema0}
\end{figure}

In the first part of this paper, the Leidenfrost system is modelled as a single component axisymmetric droplet in the Leidenfrost state, evaporating in a gaseous phase consisting of a mixture of vapour and gas, as sketched in figure \ref{fig:schema0}. 
Throughout this paper, variables and material properties associated with the liquid will be indexed by $l$. In the gaseous phase, variables and material properties of the gas, vapour and gas-vapour mixture will be indexed by $g$, $v$ and $m$ respectively.

The flow inside the droplet is modelled using the incompressible Navier-Stokes equations, as well as the energy equation:

\begin{equation}\rho_{l} \frac{D\boldsymbol{u}_{l}}{Dt} = \bnabla \bcdot \boldsymbol{\sigma}_l +\rho_{l} \boldsymbol{g}, \quad \boldsymbol{\sigma}_l = -p_l \boldsymbol{I} +\mu_{l} \left(\bnabla \boldsymbol{u}_{l} + \bnabla \boldsymbol{u}_{l}^T\right),\end{equation}
\begin{equation}
    \bnabla\bcdot \boldsymbol{u}_l=0,
\end{equation}
\begin{equation}\rho_{l} c_{p\,l} \frac{DT_{l}}{Dt}=k_{l}\nabla^2T_{l},\end{equation}

\noindent with velocity $\boldsymbol{u}$, pressure $p$, temperature $T$, density $\rho$, specific heat capacity $c_{p}$, thermal conductivity $k$, dynamic viscosity $\mu$ and the material derivative $D/Dt = \p/\p t + \boldsymbol{u}_i\bcdot\bnabla$ where $i\in\{l,m\}$ is the appropriate phase. 
The material properties $c_{p\, l}, k_l, \mu_l, \rho_l$ are taken to be constant, since they do not vary significantly with temperature. 

The governing equations in the gaseous phase are derived from the conservation of total mass, species mass (via the vapour mass fraction $w_v=1-w_g$), momentum and energy (derived from the enthalpy formulation) as in \citet{bird2006transport}:

\begin{equation}\rho_{m} \frac{D\boldsymbol{u}_{m}}{Dt} = \bnabla \bcdot \boldsymbol{\sigma}_{m} +\rho_{m} \boldsymbol{g},
\quad \boldsymbol{\sigma}_m = -p_m \boldsymbol{I} +\mu_{m} \left(\bnabla \boldsymbol{u}_{m} + \bnabla \boldsymbol{u}_{m}^T-\frac{2}{3}(\bnabla\bcdot\boldsymbol{u}_{m})\boldsymbol{I}\right),\end{equation}
\begin{equation}
    \frac{D\rho_{m}}{Dt}+\rho_{m}\bnabla\bcdot\boldsymbol{u}_{m} =0,
\end{equation} 
\begin{equation}\rho_{m}\frac{Dw_v}{Dt} = \bnabla \bcdot\left(\rho_mD_{vg}\bnabla w_v\right),\label{eqn:vap_diff}\end{equation}
\begin{equation}\rho_{m} c_{p\,m} \frac{DT_{m}}{Dt}=\bnabla\bcdot \left(k_{m} \bnabla T_{m}\right) +\frac{Dp_m}{Dt}+\rho_mD_{vg}[c_{p\,v} -c_{p\,g}]\bnabla T_m\bcdot\bnabla w_v,\label{eqn:gas_energy}\end{equation}

\noindent where the diffusion flux of vapour in equation \eqref{eqn:vap_diff} is given by Fick's law. 
In equation \eqref{eqn:gas_energy}, the last term is the enthalpy diffusion and arises due to the differing specific heats between the gas and vaporised species. 
The heating by viscous dissipation is neglected in the energy equation \eqref{eqn:gas_energy}, since it is an order of magnitude smaller than the enthalpy diffusion term. 

By combining the ideal gas law with Dalton’s law, we have the following expression for the density of the gas-vapour mixture: 

\begin{equation}\label{eqn:density}
    \rho_m = \sum_{i=\left\{v,g\right\}}\frac{p_\mathrm{atm} x_i}{R_i T_m},
\end{equation}

\noindent where $x_i$ is the mole fraction of component $i=\left\{v,g\right\}$, $R_i = R_u/M_i$ is the specific gas constant with the molar mass of each component $M_i$ and universal gas constant $R_u$. 
In equation \eqref{eqn:density}, the thermodynamic pressure of the gas-vapour mixture is replaced by atmospheric pressure $p_m \approx p_\mathrm{atm}$; a common assumption made for low Mach number flows \citep{DARU20108844}. 
This is validated by the numerical model since the maximum Mach number of the gaseous phase is $M_0^2\approx (p_m-p_\mathrm{atm})/p_\mathrm{atm}\sim 10^{-3}$.
The mole fraction is related to the mass fraction by:

\begin{equation}
    x_i = \frac{w_i/M_i}{\sum_i w_i/M_i},
\end{equation}

\noindent where the summation $\sum_i$ is understood to be over the components in the gaseous phase ${i=\left\{v,g\right\}}$.
Since there are often large variations in temperature and mass fraction in the gaseous phase, the material properties of the gas mixture vary significantly. 
Thus, we opt to consider their dependence on both variables using mixing rules as recommended in \citet{poling2000properties,bird2006transport,SOBACcompanal}:

\begin{equation}
    \mu_m = \sum_{i=\left\{v,g\right\}}\frac{x_i \mu_i}{\sum_j x_j \phi_{ij}} ,\quad \phi_{ij} = \frac{[1+(\mu_i/\mu_j)^{1/2}(M_j/M_i)^{1/4}]^2}{[8(1+M_i/M_j)]^{1/2}},
\end{equation}
\begin{equation}
    k_m = \sum_{i=\left\{v,g\right\}}\frac{x_i k_i}{\sum_j x_j A_{ij}}, \quad A_{ij}  = \varepsilon \frac{[1+(k_i/k_j)^{1/2}(M_i/M_j)^{1/4}]^2}{[8(1+M_i/M_j)]^{1/2}},
\end{equation}
\begin{equation}
    c_{p\, m} = \sum_{i=\left\{v,g\right\}} c_{p\, i} w_i,
\end{equation}

\noindent where $\varepsilon$ is a numerical constant near unity. 
Values for this constant have been reported between 0.85 and 1.1 \citep{poling2000properties}, but the exact value has a very minor effect on the results presented in the paper. 
Therefore, in this work we will use $\varepsilon=1$ since this form is self-consistent when the two species have the same properties.

The viscosity $\mu_i$ and thermal conductivity $k_i$ of the pure species are considered to depend on the temperature. 
On the considered temperature ranges, a linear fit based on experimental data is sufficient for the analysis in this paper (see appendix \ref{appA:props}).

The temperature dependence of $c_p$ in both phases is weak and so $c_p$ is taken to be constant at the saturation temperature at atmospheric pressure $T_\mathrm{sat}(p_\mathrm{atm})$, defined in the following section. 

\subsection{Boundary and interface jump conditions}\label{ssec:jump}
For convenience, we will define $j$ as the evaporative mass flux at the liquid-gaseous phase interface and variables on this interface are indexed by $I$. 
Hence, if the velocity of the interface is given by $\boldsymbol{u}_I$ and $\boldsymbol{n}$ is the outward pointing normal from the drop, then:

\begin{equation}\label{eqn:vol_flux}\rho_l(\boldsymbol{u}_l-\boldsymbol{u}_I)\bcdot \boldsymbol{n} =\rho_m(\boldsymbol{u}_m-\boldsymbol{u}_I)\bcdot \boldsymbol{n}= j.
\end{equation}

\noindent By rearranging equation \eqref{eqn:vol_flux} for velocity and integrating the governing momentum equation over a thin volume across the boundary, one obtains the following jump conditions:

\begin{equation}\label{eqn:vel_jump}
    \left[\boldsymbol{u}\bcdot\boldsymbol{n}\right]^l_m=j\left[\frac{1}{\rho}\right]^l_m,
\end{equation}
\begin{equation}\label{eqn:stress_jump}\left[-\boldsymbol{\sigma}\bcdot \boldsymbol{n}\right]^l_m = \gamma (\bnabla_S\bcdot \boldsymbol{n})\boldsymbol{n} - \bnabla_S\gamma-j^2\left[\frac{1}{\rho}\right]^l_m\boldsymbol{n},\end{equation}

\noindent where $[\,f\,]_m^l = f_l - f_m$ denotes the jump operator across the interface and $\bnabla_S$ is the surface gradient operator. 
In the majority of this work, Marangoni effects are not considered and thus the surface tension $\gamma$ is taken to be at a constant value $\gamma_0$ at $T_\mathrm{sat}(p_\mathrm{atm})$.
Conservation of the vapour species gives:

\begin{equation} \rho_mD_{vg}\bnabla w_v\bcdot\boldsymbol{n} = -j (1-w_v),\end{equation}

\noindent since the gas component is assumed to be unable to diffuse into the droplet. The jump in the energy is derived by integrating the enthalpy equation in conservative form:

\begin{equation}\label{eqn:temp_jump}\left[k\bnabla T\bcdot \boldsymbol{n}\right]_m^l = -j\left(\mathcal{L}+[c_{p\,v} -c_{p\,g}]T_m (1-w_v)\right).\end{equation}

\noindent In equation \eqref{eqn:temp_jump}, $\mathcal{L}$ is the latent heat of evaporation, which is taken to be constant due to its weak variation with temperature.
The additional term in this equation is again the contribution of enthalpy diffusion.

In addition to jump conditions, we assume that temperature and tangential velocity are continuous:

\begin{equation}
    T_l=T_m=T_I,
\end{equation}
\begin{equation}
    \left[\boldsymbol{u}\bcdot\boldsymbol{t}\right]^l_m=0.
\end{equation}

One more condition on the interface is required to close the system. 
This is the assumption of a local thermodynamic equilibrium. 
Thus, we must have that the vapour pressure is at the saturation value $p_v = p_{\mathrm{sat}}$. 
Combining this with the ideal gas law gives the following equilibrium condition:

\begin{equation}\label{eqn:equilibrium}
    \left.w_v\right|_I = \frac{p_{\mathrm{sat}}(T_I)M_v}{p_{\mathrm{sat}}(T_I)M_v+(p_{\mathrm{atm}}-p_{\mathrm{sat}}(T_I))M_g},
\end{equation}

\noindent where the same simplifying assumption of the gas-vapour mixture pressure $p_m$ as in equation \eqref{eqn:density} is justified since $(p_m-p_\mathrm{atm})/p_\mathrm{sat} \ll 1$ for the range of interface temperatures in this study. 
This has the effect of eliminating a slight over-heating underneath the droplet above the boiling temperature $T_\mathrm{sat}(p_\mathrm{atm})$ of order $0.1\,\mathrm{K}$, due to increased pressure when the vapour mass fraction is near unity.
The vapour saturation pressure is given by the following improved Antoine equation \citep{huang_simple_2018}:

\begin{equation}\label{eqn:vap_sat_p}\ln \left(p_{\mathrm{sat}}(T)/\mathrm{Pa}\right) = 34.494-\frac{4924.99}{T/\mathrm{K}-36.05}-1.57\ln{(T/\mathrm{K}-168.15)}.\end{equation}

\noindent Later in this work, we will refer to a model where the droplet sits in a pure vapour phase. 
This model is obtained when the mass fraction becomes unity everywhere, hence the equilibrium condition becomes:

\begin{equation}
    w_v \equiv 1 \quad \Leftrightarrow \quad p_\mathrm{atm} = p_{\mathrm{sat}}(T_I) \quad \Leftrightarrow \quad T_I = T_{\mathrm{sat}}(p_\mathrm{atm}).
\end{equation}

The substrate is located at $z=0$ and considered to be rigid and isothermal at temperature $T_w$. 
Hence, the variables are subject to the following boundary conditions:

\begin{equation}
\boldsymbol{u}_m=\boldsymbol{0}, \quad T_m=T_w, \quad \frac{\partial w_v}{\partial z}=0, \quad \mbox{on\ }\quad z=0.
  \label{subbc}
\end{equation}

\subsection{Quasi-stationary model}\label{ssec:quasi}
In the majority of this work, the system is assumed to be quasi-stationary (cf. §\ref{ssec:validityquasi}) which allows us to simplify our mathematical model. 
In the quasi-stationary approximation, we assume that all time derivatives in the governing equations of §\ref{ssec:gov} are zero ($\partial/\partial t=0$). 
This implies that the boundary of the droplet is stationary ($\boldsymbol{u}_I\bcdot \boldsymbol{n}=0$) and therefore the droplet volume is constant.
Assuming that $\rho_m/\rho_l\ll 1$, a consequence of equation \eqref{eqn:vol_flux} is that $\boldsymbol{u}_l\bcdot \boldsymbol{n}\approx 0$.
The mass balance across the interface is modified to reflect this:

\begin{equation}
    \rho_l\boldsymbol{u}_l\bcdot \boldsymbol{n} = 0, \quad \rho_m\boldsymbol{u}_m\bcdot \boldsymbol{n}= j.
\end{equation}

\noindent Since the ratio of gaseous mixture to liquid density is extremely small $\rho_m/\rho_l\sim \mathcal{O}(10^{-3})$, this approximation is fairly realistic. 
Further, the jump condition \eqref{eqn:stress_jump} becomes:

\begin{equation}\left[-\boldsymbol{\sigma}\bcdot \boldsymbol{n}\right]^l_m = \gamma (\bnabla_S\bcdot \boldsymbol{n})\boldsymbol{n} - \bnabla_S\gamma+\frac{j^2}{\rho_m}\boldsymbol{n}.
\end{equation}

\section{Numerical method}\label{sec:method}
The system of equations of §\ref{sec:model} is solved using a finite element method with the package \textsc{pyoomph}\footnote{Available at: \url{https://pyoomph.github.io/}}\citep{diddens_bifurcation_2024}, which is based on \textsc{oomph-lib} \citep{barth_oomph-lib_2006} and \textsc{GiNaC} \citep{bauer_introduction_2002}. 
The mesh motion in the transient model of §\ref{ssec:gov} is modelled by using an arbitrary Lagrangian–Eulerian method. 
More details are given in appendix \ref{appA:comp}.

\subsection{Far-field conditions}\label{ssec:far}
In order to model the system using finite elements, a finite computational domain must be used. 
The computational domain is given by the box $r\in[0,L]$, $z\in[0,H]$, where $L,H \approx 20 R$. In this text, $R$ without a subscript is the effective radius of the droplet, i.e., $R=(3V/4\pi)^{1/3}$, with volume $V$.
This domain size is chosen such that the near droplet velocity and temperature fields do not change significantly as the domain size increases.
We neglect the influence of lab-scale natural convection due to the heated plate and instead focus on an isolated drop with infinity-like boundary conditions.
If no stress conditions are applied on the far-field boundaries then due to the presence of gravity in the problem, there will be strong streaming down through the domain. 
This is commonly addressed by the inclusion of a no-penetration condition ($\boldsymbol{u}\bcdot\boldsymbol{n}=0$) on the boundary, with the normal pointing in the radial direction. 
Although for the behaviour near the drop this makes little difference, it significantly changes the morphology of the flow in the gaseous phase. 
Therefore, instead of a no-penetration condition, a local hydrostatic pressure distribution can be imposed using Lagrange multipliers:

\begin{equation}
\left. \begin{array}{ll}
\displaystyle  \boldsymbol{\sigma}_m\bcdot \boldsymbol{n} = -p_{\mathrm{hydro}}(z)\boldsymbol{n}
  \quad \mbox{on\ }\quad r=L,\\[8pt]
\displaystyle  \boldsymbol{\sigma}_m\bcdot \boldsymbol{n} = -p_{\mathrm{hydro}}(H)\boldsymbol{n}
  \quad \mbox{on\ }\quad z=H,\\[8pt]
  \displaystyle  \mbox{where\quad }  p_{\mathrm{hydro}}(z) = -\int_0^{z}\rho_mg\,dz.
 \end{array}\right\}
\end{equation}

\noindent The far-field boundary conditions on the temperature of the gaseous phase are simply given by zero Neumann flux:

\begin{equation}
k_m\bnabla T_m \bcdot \boldsymbol{n}= 0
  \quad \mbox{on\ }\quad r=L,\,z=H.
\end{equation}

There are two distinct regimes to consider for the boundary conditions on the gas side and top for the vapour mass fraction depending on the size of the drop. 
When the drop (and therefore domain) is large, the relative density of the different components of the gaseous mixture matters. 
In the case of water vapour, which is less dense than dry air, a plume of vapour forms above the droplet. 
Thus gas is advected through the side of the domain to replenish this, creating a strain-like flow. 
The gaseous mixture entering the domain has the ambient vapour mass fraction and therefore its vapour mass fraction will be enforced by a Dirichlet condition. 
On the top, a zero Neumann condition is employed:

\begin{equation}
\left. \begin{array}{ll}
\displaystyle   w_v=w_{\mathrm{amb}}
  \quad \mbox{on\ }\quad r=L,\\[8pt]
\displaystyle  \bnabla w_v\bcdot \boldsymbol{n}=0
  \quad \mbox{on\ }\quad z=H.
 \end{array}\right\}
\end{equation}

The presence of the Dirichlet condition raises the concern of domain size dependence. 
But as mentioned at the beginning of this section, the domain is taken large enough that there is no longer an effect of the boundary, established through testing.

When the drop is small enough, the flow away from it will be effectively radial despite the density difference in the gaseous phase. 
This is because the Grashof number $ Gr \sim \rho_m^2gR^3/\mu_m^2\max\left(\Delta R_m/R_m,\Delta T_m/T_m\right)$ for drops $R\lesssim 100\,\upmu\mathrm{m}$ is $Gr\lesssim\mathcal{O}(10^{-3})$, since the variations in the gas constant and temperature are no more than $\mathcal{O}(1)$.
In this regime, the Péclet number for vapour transport ($\Pen_m =  U_m R/D_{vg}$) in the gas mixture is small due to the small droplet size. 
In the far field, the drop may be approximately considered to be a point source of vapour, with the vapour mass fraction solving the Laplace equation $\nabla^2w_v=0$.
The far-field behaviour of the Green's function $w_v \sim 1/|\boldsymbol{x}|$ is captured to first order by the Robin condition as employed in \citet{diddens_detailed_2017}:

\begin{equation}
\bnabla w_v\bcdot \boldsymbol{n} =  (w_v-w_\mathrm{amb})\frac{\boldsymbol{x}\bcdot\boldsymbol{n}}{|\boldsymbol{x}|^2}
  \quad \mbox{on\ }\quad r=L,\,z=H.
\end{equation}

\noindent Since the Péclet number in the gaseous mixture is small, vapour is not quickly advected out of the domain.
Hence, if a zero Neumann condition is erroneously used here, the vapour is unable to vacate the domain which unrealistically  fills with vapour.

\subsection{Example simulation}\label{ssec:exsim}
\begin{figure}
  \centerline{\includegraphics[width=\textwidth]{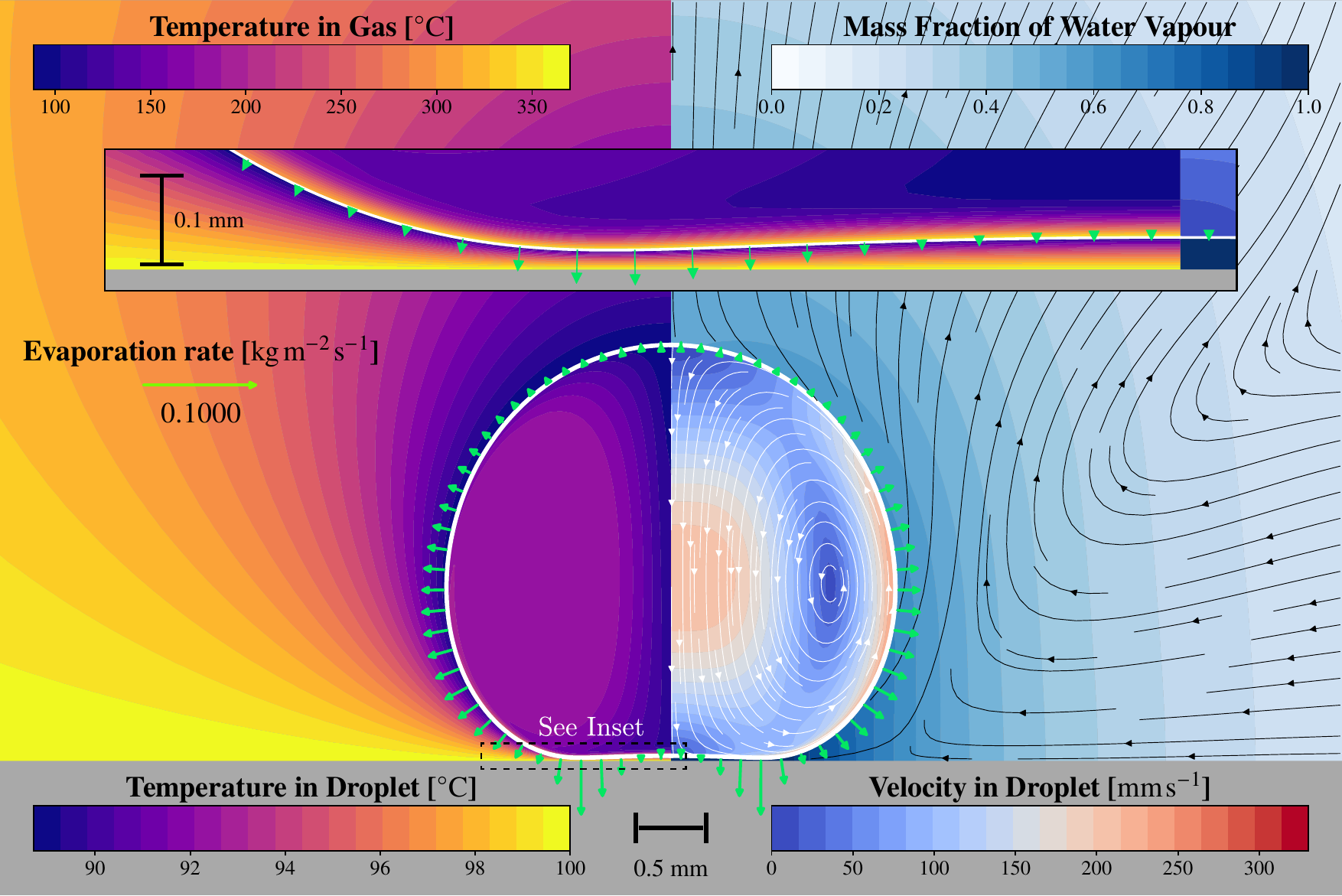}}
  \caption{Snapshot from simulation of a quasi-stationary pure water droplet with $R=1.59\,\mathrm{mm}$ above a plate at $T_w = 370 \,^\circ\mathrm{C}$. The temperature field in the droplet and gas phase are plotted on the left (Note: for visibility purposes each domain has a different colour bar which share the same minimum temperature, since temperature variations in the drop $\sim10\,\mathrm{K}$ are far smaller than that of the gas $\sim200\,\mathrm{K}$). The velocity field in the droplet; mass fraction field in the gas phase; as well as streamlines in both phases are plotted on the right. The inset shows a blow-up of the vapour layer and the thermal boundary layer in the drop.}
\label{fig:snap0}
\end{figure}

In figure \ref{fig:snap0}, a simulation of a quasi-stationary water droplet of size $R=1.59\,\mathrm{mm}$ is shown. 
The droplet shows the expected characteristics of Leidenfrost droplets of this size, namely a quasi-spherical shape with a flattened bottom and a thin vapour layer of order $30 \,\upmu\mathrm{m}$. 
We also see that the mass fraction of water vapour in the flow in the vapour layer is close to 100\% and the ejected vapour meets the oncoming gas at a stagnation point creating a vertical plume of vapour.
There is a significant effect of vapour cooling at the top of the droplet of order $\sim 10\,\mathrm{K}$.
The cooler liquid is convected down the $z$ axis and around the interface by strong flows driven by shear with the vapour layer.
Note that in this simulation the thermal Marangoni effect is disregarded.

\section{Droplet shapes}\label{sec:shapes}
The mathematical model outlined in the previous sections is used to investigate the quasi-static Leidenfrost system for a pure water droplet. 
The different fields in this system are strongly coupled.
Thus, it is difficult to separate the roles of ambient humidity and the flow in the drop. 
To cope with this, we will first consider less realistic models with neglected flow in the drop and/or a pure vapour phase.
These models are summarised in table \ref{tab:models}, where the first model is the most realistic model and subsequent models have the additional assumptions mentioned.
In order to eliminate the internal circulation in the drop for the relevant models, the viscosity of the liquid is increased by a factor of $10^4$.
This is sufficiently large so that upon further increase there is negligible change in the global evaporation rate for droplet sizes covered in the analysis, thus decoupling its effect.

\begin{table}
  \begin{center}
\def~{\hphantom{0}}
  \begin{tabular}{llcc}
      Model &Description  & $w_\mathrm{amb}$  & $\mu_l/\mu_w$ \\[3pt]
       MGV & Mixed gas-vapour gaseous phase model   & 0 & 1 \\
       PV & Pure vapour phase model   & 1 & 1 \\
       MGV-NC & Mixed gas-vapour gaseous phase model without internal drop circulation & 0 & $10^4$ \\
       PV-NC & Pure vapour phase model without internal drop circulation & 1 & $10^4$ \\
  \end{tabular}
  \caption{Overview of the different models used in this work.
  NC in the model names stands for neglected (internal drop) circulation.
  $w_\mathrm{amb}$ is the ambient vapour mass fraction of the gaseous phase and $\mu_l$ is the viscosity of the liquid, where $\mu_w$ is the viscosity of water.}
  \label{tab:models}
  \end{center}
\end{table}

To permit comparison of our model with experimental analysis of water droplets, we ran simulations with parameters outlined in table \ref{tab:mat_props} in appendix \ref{appA:props}. 
Parameters such as the substrate temperature are changed in order to facilitate comparison with different experimental data. 
This is clearly indicated whenever it is the case. 

\begin{figure}
  \centerline{\includegraphics[width=\textwidth]{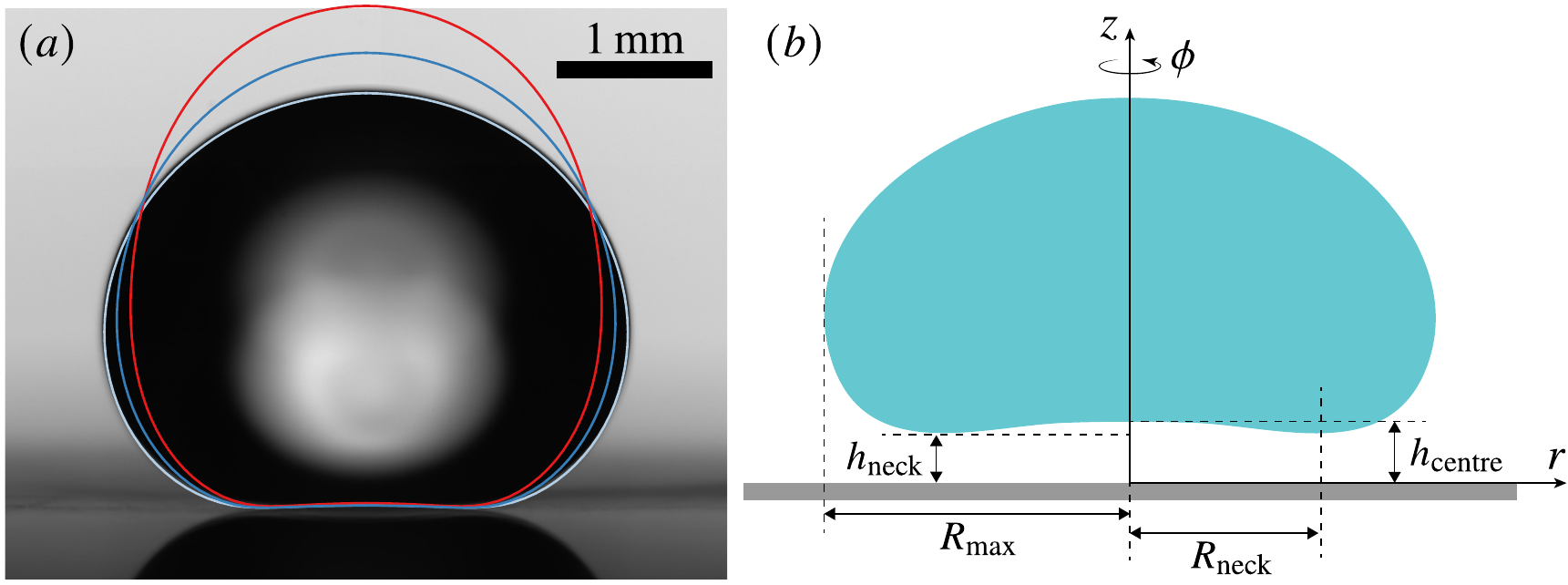}}
  \caption{$(a)$ Image of a water drop in the Leidenfrost state with $R=1.59\,\mathrm{mm}$ adapted with permission from \citet{burton_geometry_2012} (Copyrighted by American Physical Society.), with droplet shapes from the numerical models superposed. 
  From top to bottom are the simulations with a pure vapour gaseous phase with drop circulation (PV in red), mixed gas-vapour phase with drop circulation (MGV in blue) and finally, both models with neglected drop circulation (MGV-NC in light blue, PV-NC in peach) which to the eye overlap perfectly at this scale.
  $(b)$ Schematic of a Leidenfrost drop with key vapour layer metrics pictured, these are: $R_\mathrm{max}$, the maximum droplet radius; $R_\mathrm{neck}$, the radius of the neck region which is measured at the minimum vapour thickness, $h_\mathrm{neck}$; $h_\mathrm{centre}$, the height at $r=0$.}
\label{fig:burton}
\end{figure}

In figure \ref{fig:burton}$(a)$, droplet profiles from the four numerical models are plotted over a photograph from the experiments of \citet{burton_geometry_2012}. 
For this drop size of $R=1.59\,\mathrm{mm}$, we see that the models where the internal drop flow is completely neglected surprisingly and unexpectedly agree best with the shape from experiments, whereas the more realistic models show more prolate shapes. 
We also observe that considering a mixed gas-vapour phase predicts overall drop shapes closer to those of experiments. 

\begin{figure}
  \centerline{\includegraphics[width=\textwidth]{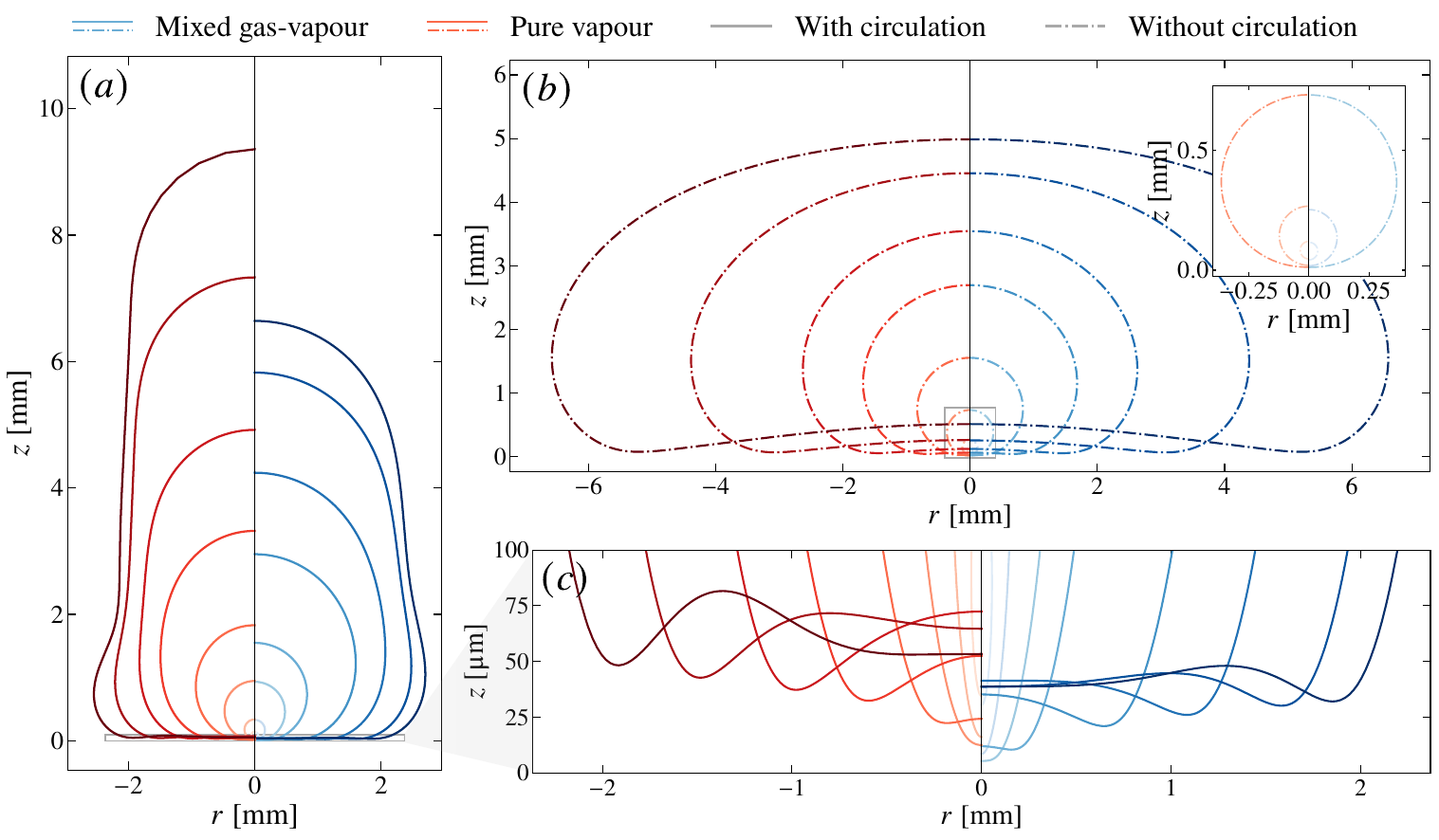}}
  \caption{Quasi-stationary Leidenfrost water drops shapes from numerical simulations with substrate temperature $T_w=370\,^\circ\mathrm{C}$. $(a)$ Mixed and pure vapour gas phase models (in blue and red respectively) at viscosity $\mu_l=\mu_w$ with drop sizes varying from $R\approx0.05\,\mathrm{mm}$ to $R\approx3\,\mathrm{mm}$. $(b)$ The same models as in $(a)$ but with $\mu_l=10000\mu_w$ from $R\approx0.05\,\mathrm{mm}$ to $R\approx5\,\mathrm{mm}$ . $(c)$ Blow-up of the thin vapour film in grey rectangle in $(a)$.}
\label{fig:shapes}
\end{figure}

Plotted in figure \ref{fig:shapes} are the drop shapes for each of the different models varying from drop radii $R\approx 0.05\,\mathrm{mm}\rightarrow5\,\mathrm{mm}$. 
In the cases of eliminated drop circulation, we see the familiar Leidenfrost drop shapes and the transition from large puddles to dimpled quasi-spherical drops as shown in experiments from \citet{burton_geometry_2012} and lubrication theory of \citet{sobac_leidenfrost_2014}, as well as a transition to a dimple-less regime \citep{sobac_erratum_2021}. 
However, if now the viscosity is reduced to that of water for large drops with $R\gtrsim\ell_c$,  we observe very prolate drop shapes as in figure \ref{fig:shapes}$(a)$. 
These shapes are accompanied by the formation of a `wimple' vapour layer structure as reported in the coupled lubrication Navier-Stokes computational model of \citet{Chakraborty_Chubynsky_Sprittles_2022}. 
A hypothesis for the cause of these prolate shapes will be discussed in §\ref{sec:stab}.

\begin{figure}
  \centerline{\includegraphics[width=\textwidth]{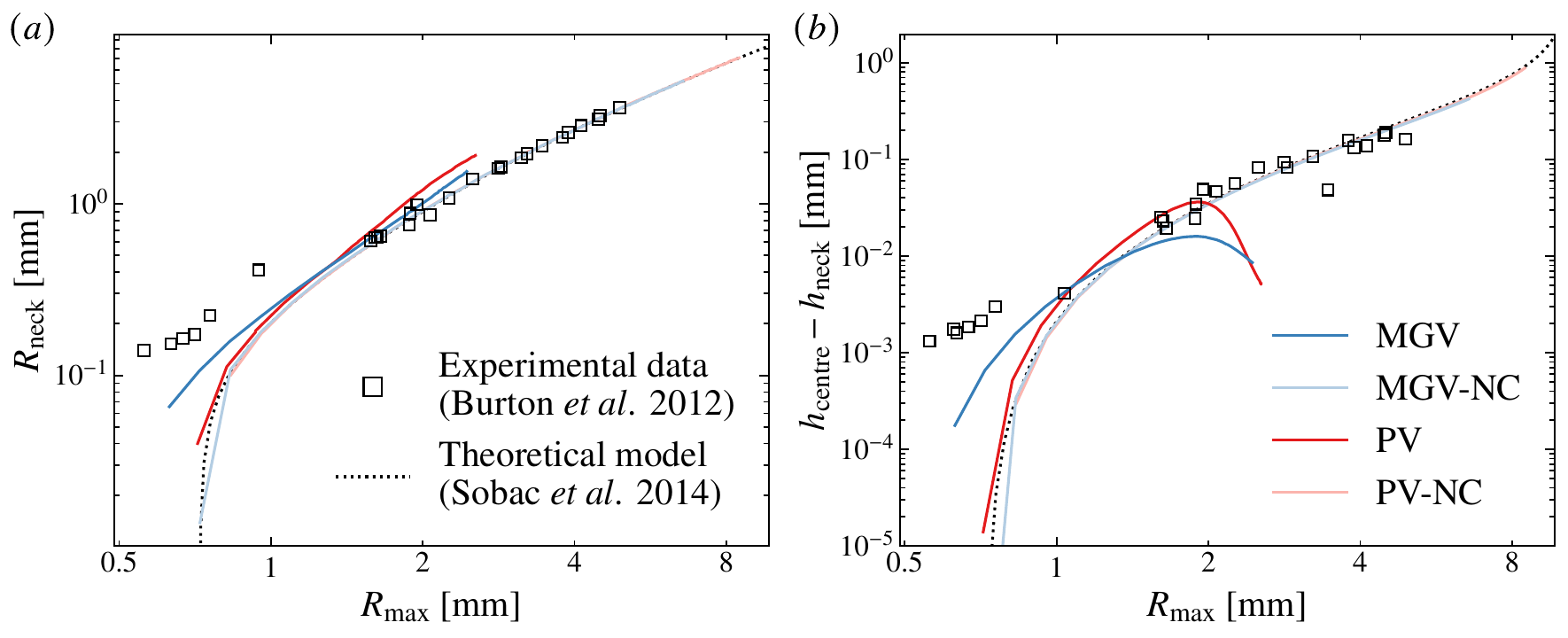}}
  \caption{($a$) $R_\mathrm{neck}$ and ($b$) $h_\mathrm{centre}-h_\mathrm{neck}$, against $R_\mathrm{max}$ calculated from the numerical models with substrate temperature $T_w=370^\circ\mathrm{C}$ for water droplets. Also plotted are the experimental results from \citet{burton_geometry_2012} and the theoretical results of the model of \citet{sobac_leidenfrost_2014}.}
\label{fig:rn}
\end{figure}

As done in previous works, to permit quantitative comparison between the model and experiments, we define geometric characteristics of the vapour layer as depicted in the schematic in figure \ref{fig:burton}$(b)$. 
In figure \ref{fig:rn}, $R_\mathrm{neck}$ and $h_\mathrm{centre}-h_\mathrm{neck}$ are plotted for each of the models against $R_\mathrm{max}\gtrsim 0.5\, \mathrm{mm}$. 
The reason for this cut-off is due to the drops transitioning from a dimple structure to a quasi-spherical dimple-less regime. 
The balance between the lubrication pressure in the thin vapour layer and the Laplace pressure of the drop gives the length scale at which this occurs, as was derived in the work of \citet{POMEAU2012867}. 
This is given by:

\begin{equation}
    \ell_i = \left(\frac{\mu_v k_v \gamma^2\Delta T }{\rho_l^3 \rho_v g^3 \mathcal{L}}\right)^{1/7}\approx 0.4\,\mathrm{mm}.
\end{equation}

The shapes of the high viscosity liquid models expectedly agree well with the theoretical lubrication model of \citet{sobac_leidenfrost_2014}.
In the large radius limit of figure \ref{fig:rn}, the `wimple' structure in the models considering drop circulation can be recognised by the deviation from the experimental and theoretical data. 

\begin{figure}
  \centerline{\includegraphics[width=\textwidth]{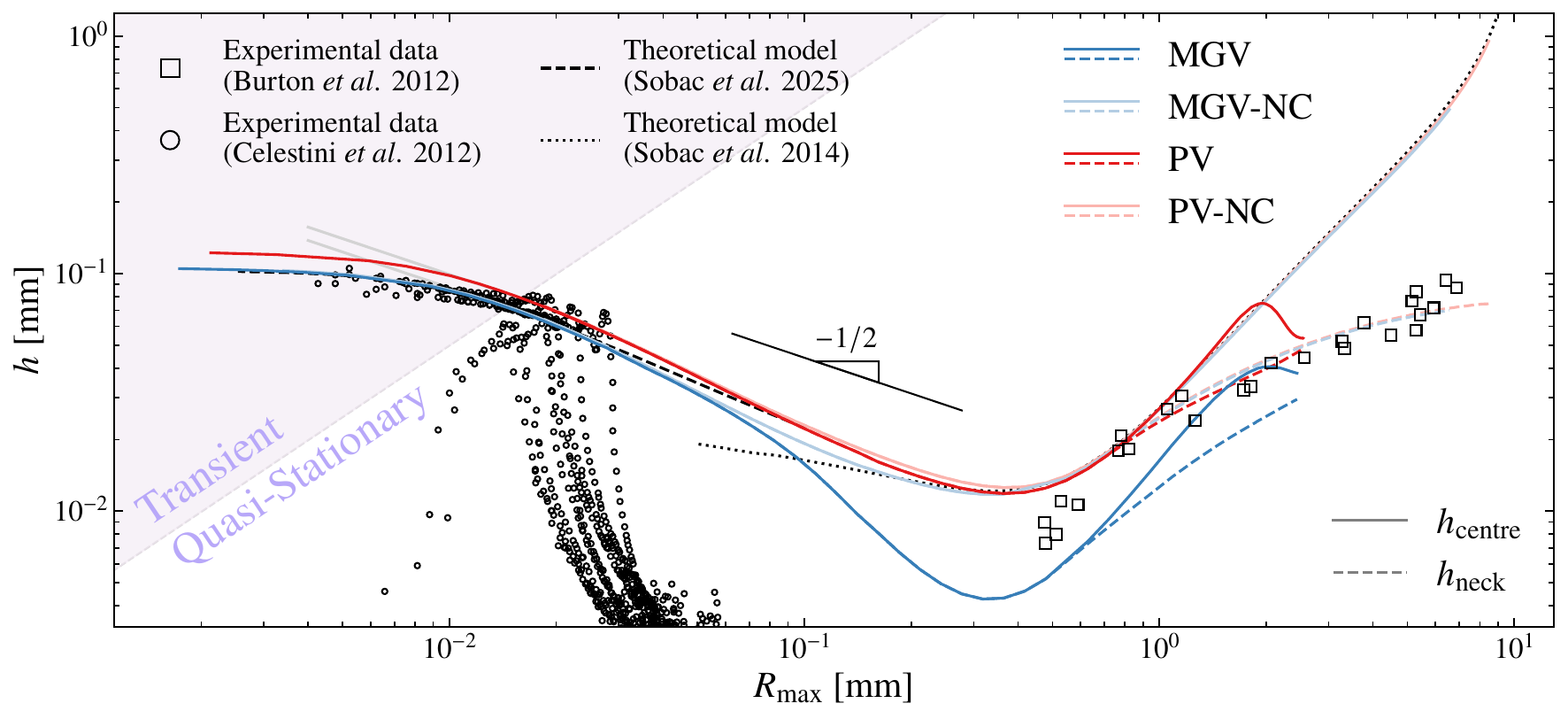}}
  \caption{$h_\mathrm{centre}$ and $h_\mathrm{neck}$ against $R_\mathrm{max}$ from the numerical models with experimental data for $h_\mathrm{neck}$ from \citet{burton_geometry_2012} with $T_w=370^\circ\mathrm{C}$. The grey lines in the part of the figure marked transient indicate quasi-stationary simulations in this region. The data for take-off Leidenfrost droplets from \citet{celestini2012} are at the slightly higher $T_w=400^\circ\mathrm{C}$, but they are included here anyhow for rough comparison. The results from the theoretical models from \citet{sobac_leidenfrost_2014} and \citet{Sobac_Rednikov_Colinet_2025} for large and small Leidenfrost drops are also included.}
\label{fig:h}
\end{figure}

A more revealing metric to analyse is the droplet separation from the substrate, since this remains defined across the entire lifetime of the Leidenfrost droplet. 
In figure \ref{fig:h}, both  $h_\mathrm{neck}$ and $h_\mathrm{centre}$ are plotted over four decades of droplet radius. 

On the right side of figure \ref{fig:h}, for large Leidenfrost droplets, as before we observe good agreement of the models neglecting circulation with the lubrication model of \citet{sobac_leidenfrost_2014} and for $h_\mathrm{neck}$ with the experimental data from \citet{burton_geometry_2012}.
The ambient humidity does not affect the results of these models, as the relevant physics occurs within the thin vapour layer which essentially consists of pure vapour.
Where the solid and dashed curves coalesce is the point where the droplets transition to a dimple-less regime.
The unphysical `wimple' structure can also be seen for the models with circulation when $h_\mathrm{centre}$ begins to decrease for large drops.

On the left side of the figure, the results from the theoretical model of \citet{Sobac_Rednikov_Colinet_2025}, which considers a spherical droplet evaporating in a field of pure vapour, are given by a dotted black line. 
Although there is nice agreement overall, the final take-off height for the pure models in this paper deviates from that for this theoretical model. 
This is because the model of \citet{Sobac_Rednikov_Colinet_2025} uses material properties in the gaseous phase evaluated at the average temperature between the drop and the substrate $(T_w+T_d)/2$, a so-called 1/2 rule. 
In this limit, as the Leidenfrost drop takes off from the substrate, it approaches the well-studied case of isolated droplet evaporation. 
Here it is known that the 1/2 rule is less accurate than the 1/3 rule, where the material properties are evaluated one-third of the way (by linear interpolation) between the temperature of the droplet and the far field \citep{HUBBARD19751003}. 
If the material properties of the pure vapour gaseous phase of the model presented in this paper are taken to be constant at the 1/2 rule, there is full agreement with the theoretical curve of \citet{Sobac_Rednikov_Colinet_2025}.

The experimental results from the take-off experiments of \citet{celestini2012} are also plotted. 
The take-off occurs roughly at the take-off length scale $\ell_*$, which is defined as when the droplet leaves the lubrication regime (i.e., $h\sim R$). This is derived in \citet{celestini2012}:

\begin{equation}\label{eqn:ell_star}
    \ell_* = \left(\frac{\mu_m k_m \Delta T}{\rho_l \rho_m g \mathcal{L}}\right)^{1/3}\approx 30 \,\upmu\mathrm{m}.
\end{equation}

It is important to note that these experimental results are at a slightly different substrate temperature of $T_w=400\,^\circ\mathrm{C}$ as compared to the $T_w=370\,^\circ\mathrm{C}$ of the rest of the figure. 
Since the relative temperature change is small ($\Delta T_w/T_w\approx0.05$), we expect them to be reasonable points of comparison. 
Indeed, it can be seen that the upper experimental points agree well with the curves for drop height, while the lower points are deemed to be the result of effects not captured in this model (cf. § \ref{ssec:ma}).

In this limit of small Leidenfrost drops, the models considering a mixed gas-vapour phase sit systematically lower than their pure vapour counterparts. 
This is not solely to do with differing temperature gradients due to evaporative cooling (discussed in §\ref{sec:humcirc}), but also since all material properties will change significantly with temperature and composition.
Thus, the characteristic length scale for the drop height will differ between the models with different ambient humidity.

The shaded transient region on the left-side of the figure demarcates roughly where the quasi-stationary model can no longer be applied. 
This was identified in the theoretical work of \citet{Sobac_Rednikov_Colinet_2025} on spherical take-off Leidenfrost droplets plotted on the figure. 
The drop can no longer be considered quasi-stationary because the Stokes' drag becomes of similar order to the evaporative thrust. 
By balancing the viscous drag and the evaporative force one obtains $(h/R)^3\sim \rho_l/\rho_m$.
The dividing line was therefore chosen to be $h/R= 0.5 (\rho_l/\rho_m)^{1/3}$, acting as a rough boundary of this domain. 
Indeed, if the system is wrongly assumed to be quasi-stationary, drops in this region soar to infinite height with $h\propto R^{-1/2}$ in accordance with results from \citet{Sobac_Rednikov_Colinet_2025}. 
This is shown in the figure by quasi-stationary solutions for both the pure and mixed vapour-gas phase in grey. 
The coloured curves are obtained by running transient simulations from the quasi-stationary regime into the transient regime. 
If a transient simulation is initialised at a height different from the equilibrium in the quasi-stationary regime, solutions will decay quickly onto the equilibrium, much like for a damped oscillator.

In a rather nice way, this numerical model smoothly connects the two separate regimes of the theoretical works of \citet{sobac_leidenfrost_2014} and \citet{Sobac_Rednikov_Colinet_2025}. 
We can see that the large Leidenfrost drop theoretical model begins to deviate from the models presented in this study, due to the smallness parameter in lubrication theory becoming relatively large. 
Similarly, from the other direction the sphericity assumption becomes inappropriate as we approach $R\sim\ell_i$. The intermediate scaling of $h\sim R^{-1/2}$ in the take-off regime $R\sim \ell_*$ is recovered in our model and is well known since \citet{POMEAU2012867} and \citet{celestini2012}. 

Perhaps what is most striking in figure \ref{fig:h} is the significantly different behaviour of the mixed gas-vapour model with internal circulation at the bottom of the figure. 
As the drop size increases, at around $R_\mathrm{max}= 50 \,\mathrm{\upmu m}$ it departs from the other models, predicting much thinner vapour layers.
It appears to more closely capture the behaviour of the leftmost experimental results of \citet{burton_geometry_2012}.
This will be discussed further in later sections.

\section{Azimuthal stability}\label{sec:stab}

\begin{figure}
  \centerline{\includegraphics[scale=0.35]{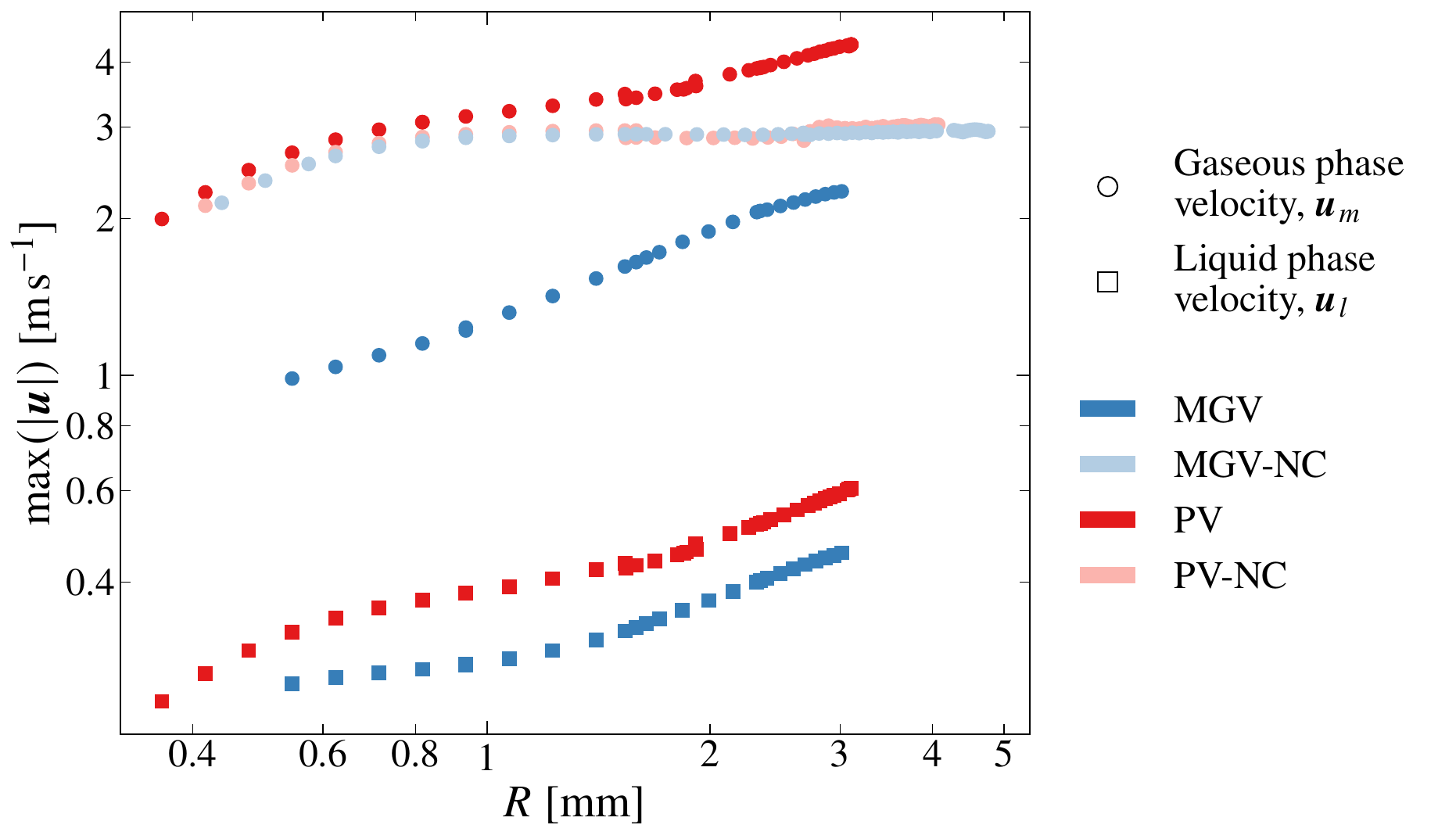}}
  \caption{Maximum velocity magnitude in the liquid and gaseous phases for large Leidenfrost drop volumes and temperature of the substrate $T_w=370\,^\circ\mathrm{C}$. 
  The maximum drop velocities in the axisymmetric model are considerably higher than those seen in experiments.}
\label{fig:vel}
\end{figure}

The most unexpected result of the previous section was that the more accurate mathematical models (where drop circulation is considered) for large droplets predicted prolate drop shapes which are not observed experimentally. 
These shapes are a consequence of over-predicted internal circulation velocities driving upwards, augmenting the drop shape.
This can be seen from the maximum velocity magnitudes in both phases plotted in figure \ref{fig:vel}. 
In the gaseous phase, this velocity is attained in the neck region and similarly the maximum velocity in the drop occurs on the other side of the interface driven by the gas shear.

The internal droplet circulation in experimental Leidenfrost drops has been visualised and quantified by particle image velocimetry (PIV) measurements by \citet{bouillant_leidenfrost_2018}. 
They found the internal drop velocities to be roughly of the order of a few $\mathrm{cm}\,\mathrm{s}^{-1}$, an order of magnitude lower than our models with circulation considered. 
This leads one to question what physical effects are not being captured in the present model which cause this dramatic effect.
A significant simplifying assumption in our numerical model is that of axisymmetry.
The PIV measurements of \citet{bouillant_leidenfrost_2018} show that millimetric droplets clearly exhibit axisymmetry breaking flows, which, as the droplet evaporates, transition through successive azimuthal modes. 
Such non-axisymmetric flows may alter the shape of the droplet and are likely to have different characteristic velocities due to azimuthal averaging.
Thus, if the quasi-stationary axisymmetric state is unstable to azimuthal perturbations, it is possible that such a shape would not be observed in experiments. 
The flow is expected to become axisymmetric when the droplet becomes small enough due to the stabilising nature of the small geometry. 
We therefore seek to find the onset of instability for our most complete model.

The azimuthal stability analysis is conducted by solving the generalised eigenproblem for the system, as outlined in \citet{diddens_bifurcation_2024}. 
All scalars ($c$) and vectors ($\boldsymbol{v}$) of the axisymmetric base state are perturbed as follows:

\begin{align}
    c = c_\mathrm{axi}(r,z) + \varepsilon c_m(r,z) e^{im\phi} e^{\lambda t}, \quad \boldsymbol{v} = \boldsymbol{v}_\mathrm{axi}(r,z) + \varepsilon \boldsymbol{v}_m(r,z) e^{im\phi} e^{\lambda t}, 
\end{align}

\noindent where $\boldsymbol{v}_m$ may also have a component in the $\phi$ direction. 
The eigenvalues $\lambda$ are then acquired using the shift-inverted Arnoldi method \citep{arnoldi,saad}. 

\begin{figure}
  \centerline{\includegraphics[width=\textwidth]{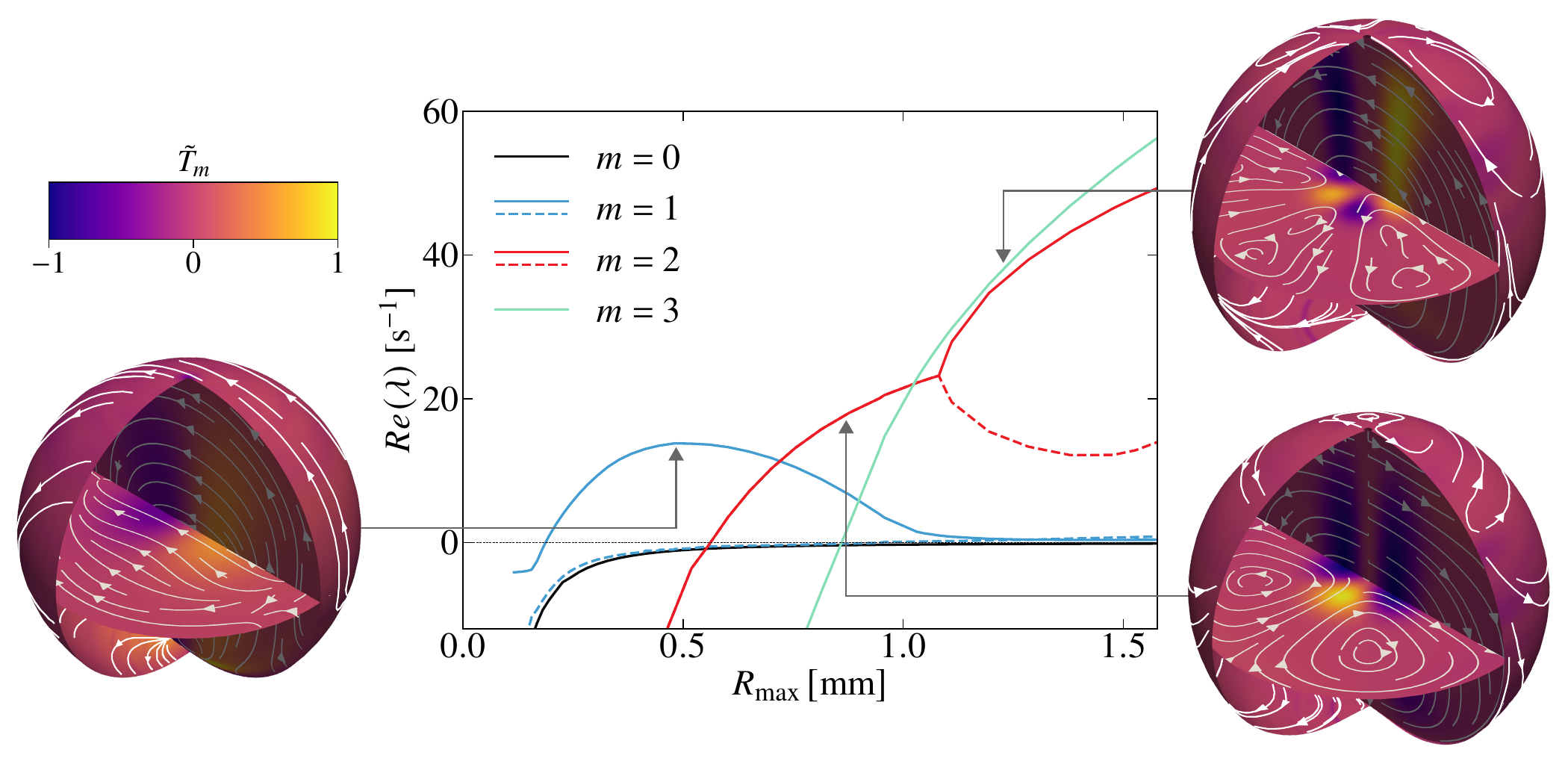}}
  \caption{Growth rates of the unstable (or most unstable) eigenmodes for azimuthal wavenumbers $m=0,1,2,3$ for the most complete model of mixed gaseous phase and internal drop circulation at substrate temperature $T_w=370\,^\circ\mathrm{C}$. 
  These are plotted with maximum drop extent of the axisymmetric base state in the range $0.1\,\mathrm{mm}\lesssim R_\mathrm{max}\lesssim1.6\,\mathrm{mm}$. 
  The adjacent plots are the most unstable eigenmodes in each regime visualised in 3D. 
  The colourmap indicates the normalised temperature eigenmodes and streamlines indicating the velocity in each respective surface.}
\label{fig:stability}
\end{figure}

In figure \ref{fig:stability}, we can see the growth rates of the most unstable eigenmodes for the azimuthal wavenumbers $m=0,\,1,\,2,\,3$, as well as visualisations of the corresponding 3D eigenmodes for the temperature field.

We observe a cascade of dominant modes as the drop evaporates from $m=3\rightarrow2\rightarrow 1$.
This cascade has been observed experimentally in the PIV measurements by \citet{bouillant_leidenfrost_2018} and also numerically by \citet{yim_leidenfrost_2022}, who used azimuthal stability analysis to perturb non-wetting drop base states with a temperature distribution from an experimental result imposed across the surface of the droplet. 
Although qualitatively similar, the radii at which transitions between different modes occur in this work are quite different from those in the study of \citet{yim_leidenfrost_2022} and the experimental results of \citet{bouillant_leidenfrost_2018}. 
Presumably, this is because the model in this paper is much more complete than that of \citet{yim_leidenfrost_2022}, where the shear from the vapour layer is omitted and the effect of the outer gaseous phase is modelled by an imposed temperature distribution. 
However, \citet{yim_leidenfrost_2022} include thermal Marangoni effects (unlike the analysis in this paper thus far, cf. § \ref{ssec:ma}) via an effective Marangoni number to match the order of magnitude internal velocities from experiments.
This drives a Marangoni flow in the same direction as shear from the gas since the drop is cooler at the top. 
It is possible that the lack of vapour shear is compensated by the Marangoni-driven flows giving qualitative agreement.

As expected, once the drop becomes smaller than $R_{\mathrm{max}} = R \approx 0.2\,\mathrm{mm}$, it becomes stable against the non-axisymmetric distortions. 
This threshold value defines the stable axisymmetric bound.
The $m=1$ eigenmode is a special case and represents the structure of the solid-like rolling of the Leidenfrost wheels of \citet{bouillant_leidenfrost_2018}.
The critical onset radius of azimuthal instability of $m=1$ has up to now never been reported numerically, or experimentally.
The bound of $R \approx 0.2\,\mathrm{mm}$ lies near the minimum of the gap width $h$ of the most complete model in figure \ref{fig:h}.
This result gives us some confidence that the predicted lower droplet heights of the most complete model versus the simplified models reflect a genuine effect, which seemingly may better capture behaviour of the experimental data of \citet{burton_geometry_2012}.
Thus, it is expected that conclusions drawn on the underlying mechanism and its implications are valid even beyond this stability limit in the following sections.

\section{Ambient humidity and drop circulation}\label{sec:humcirc}
\begin{figure}
  \centerline{\includegraphics[width=\textwidth]{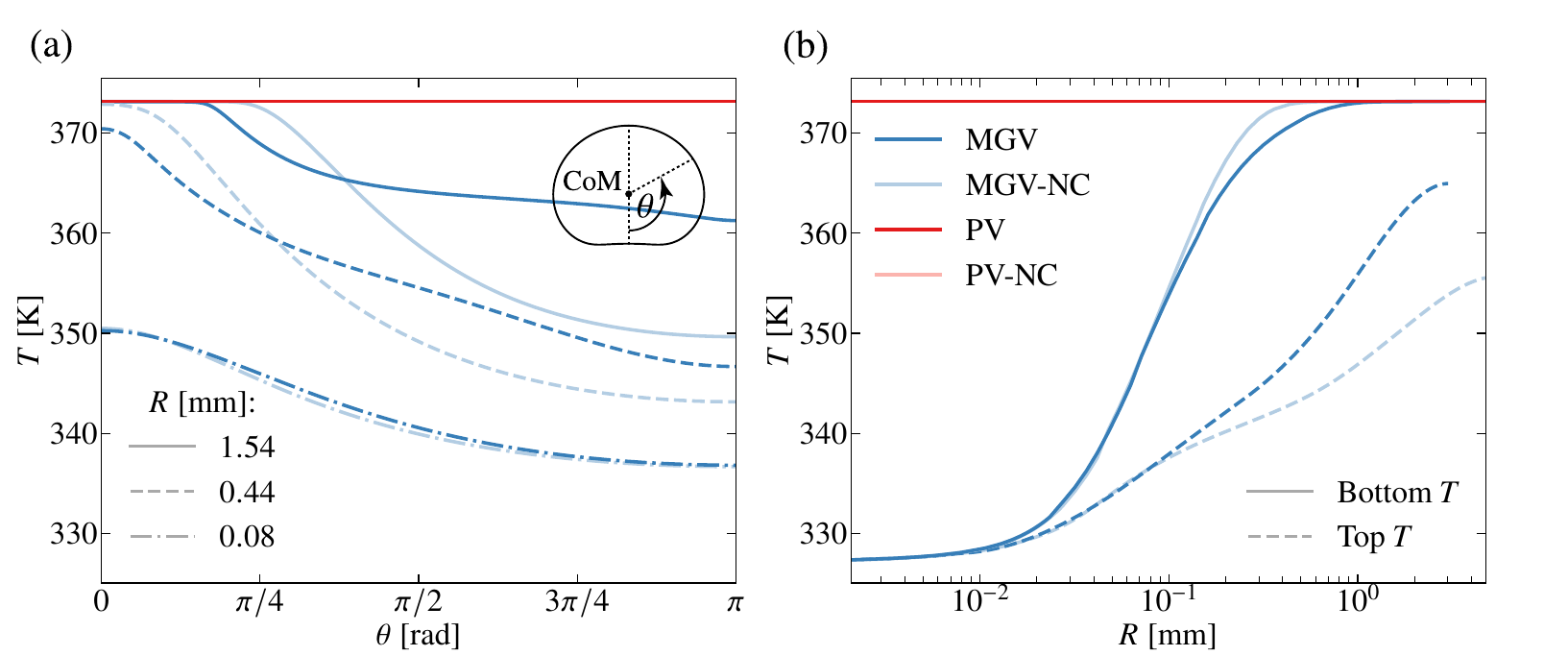}}
  \caption{$(a)$ Temperature around the surface of the drop for various drop sizes at substrate temperature $T_w=370\,^\circ\mathrm{C}$.
  The polar angle $\theta$ is depicted in the inset diagram and defined as the angle between the $z$ axis and the line through the centre of mass and a point on the droplet surface, where 0 is at the drop of the bottom.
  $(b)$ Temperature at the bottom and top of the drop at $r=0$ against volume of the droplet. 
  In both $(a)$ and $(b)$ the models with a pure vapour gaseous phase overlap perfectly as both are isothermal.
  }
\label{fig:all_temps}
\end{figure}

The equilibrium condition \eqref{eqn:equilibrium} at the interface mediates the effect of the ambient humidity on the drop temperature.
From this condition, we see that if the mass fraction of vapour is identically $w_v\equiv1$, then the temperature at the interface of the drop is at the boiling temperature (i.e., $T=100\,^\circ \mathrm{C}$ for water at atmospheric pressure). 
Conversely, if there is a non-unity mass fraction of vapour, then the interface can cool below the saturation temperature at atmospheric pressure. 
Indeed, this is reflected by the models in figure \ref{fig:all_temps}. 

In figure \ref{fig:all_temps}$(a)$, we can see typical temperature profiles along the surface of the drop for various drop sizes. 
For drops larger than $R\gtrsim\ell_i$, when a thin vapour layer is present, the underside of the droplet is at the boiling temperature, while the top of the droplet in the mixed models is evaporatively cooled by order $\sim 10\,\mathrm{K}$. 
Evidence of such evaporative cooling of the droplet is seen in supplementary figure 9 of \citet{bouillant_leidenfrost_2018}, where we can also observe thin thermal boundary layers at the bottom of the droplet. 
In the models with a mixed gas-vapour phase, internal drop circulation reduces temperature differences in the liquid phase due to enhanced mixing.
As the drop size decreases, the entire drop evaporates at temperatures below the saturation temperature at atmospheric pressure, as displayed in figure \ref{fig:all_temps}$(b)$.
The drop continues to evaporate at these low temperatures because latent heat is still supplied to the interface by conduction through the gas, while vapour is diffused away. 
Consequently, the drop must evaporate to maintain thermodynamic equilibrium.
This degree of cooling throughout the entire drop is consistent with the theoretical work of \citet{SOBACcompanal} on isolated drop evaporation.
The radius below which the effect of internal circulation becomes negligible is evident from figure \ref{fig:all_temps}$(b)$.
This occurs when the liquid Péclet number $\Pen_l$ becomes order unity, which can also be seen when the blue curves coalesce in figure \ref{fig:h}.

\begin{figure}
  \centerline{\includegraphics[scale=0.33]{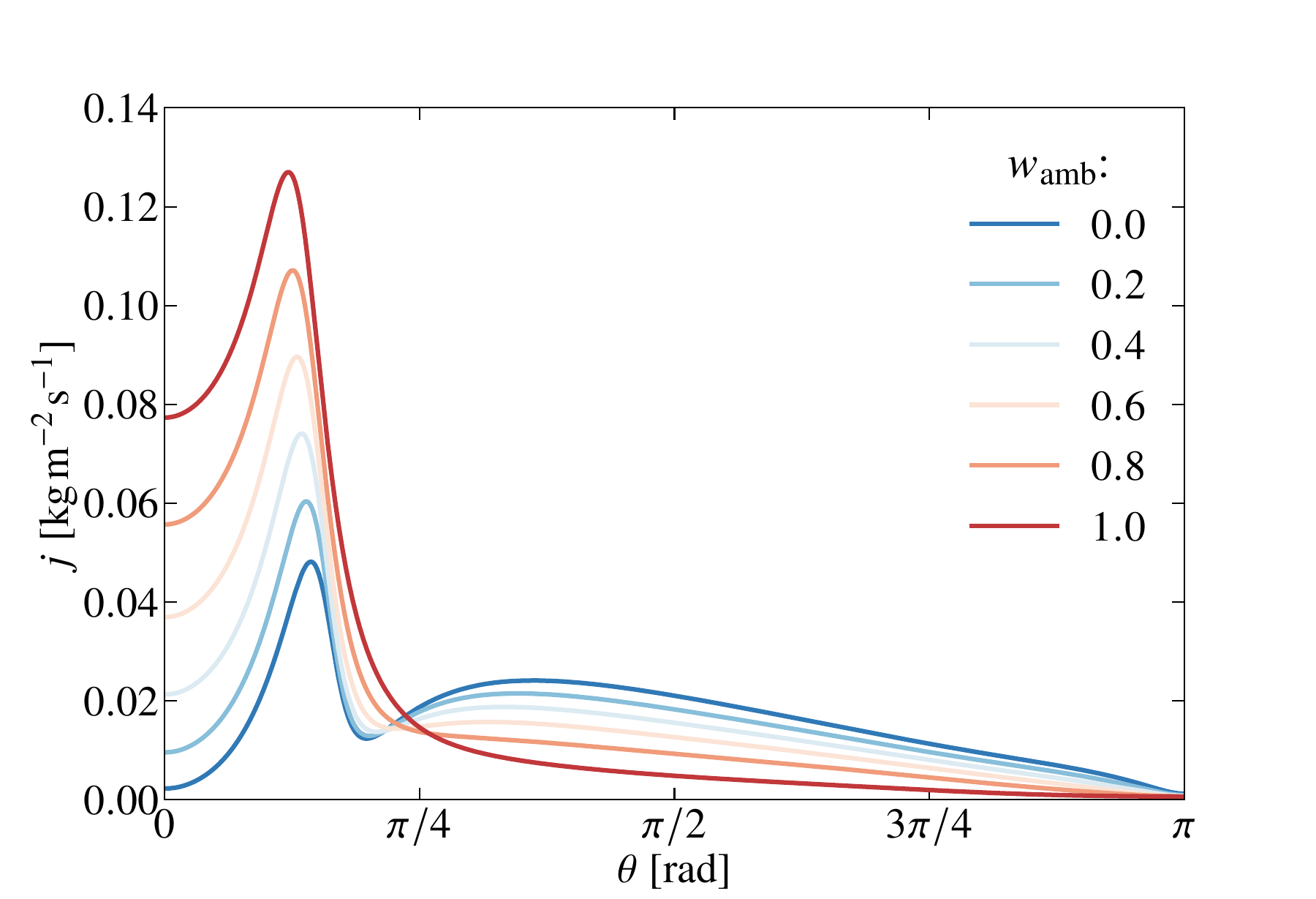}}
  \caption{Evaporation flux $j$ against $\theta$ (as defined in figure \ref{fig:all_temps}) at the liquid-gas interface for changing ambient humidity for a drop of size $R=1.59\,\mathrm{mm}$ and substrate temperature $T_w=370\,^\circ\mathrm{C}$.}
\label{fig:m2p}
\end{figure}

Figure \ref{fig:m2p} shows the evaporative flux $j$, as introduced in §\ref{sec:model}, given as a function of the polar angle $\theta$ for a Leidenfrost droplet with drop circulation where the far-field ambient humidity is increased. 
As the ambient humidity increases, the evaporation rate at the bottom of the droplet is significantly increased, while the evaporation rate outside is decreased.
The main difference between the model with $w_\mathrm{amb}=0$ and $w_\mathrm{amb}=1$ is the evaporative cooling outside the film layer due to the presence of a mixed gas-vapour phase.
However, it is only when internal droplet circulation is considered in addition to this cooling that this information is propagated to the bottom of the droplet. 
This indicates a strong coupled role of the internal circulation and ambient humidity in quasi-stationary models.

\begin{figure}
  \centerline{\includegraphics[width=\textwidth]{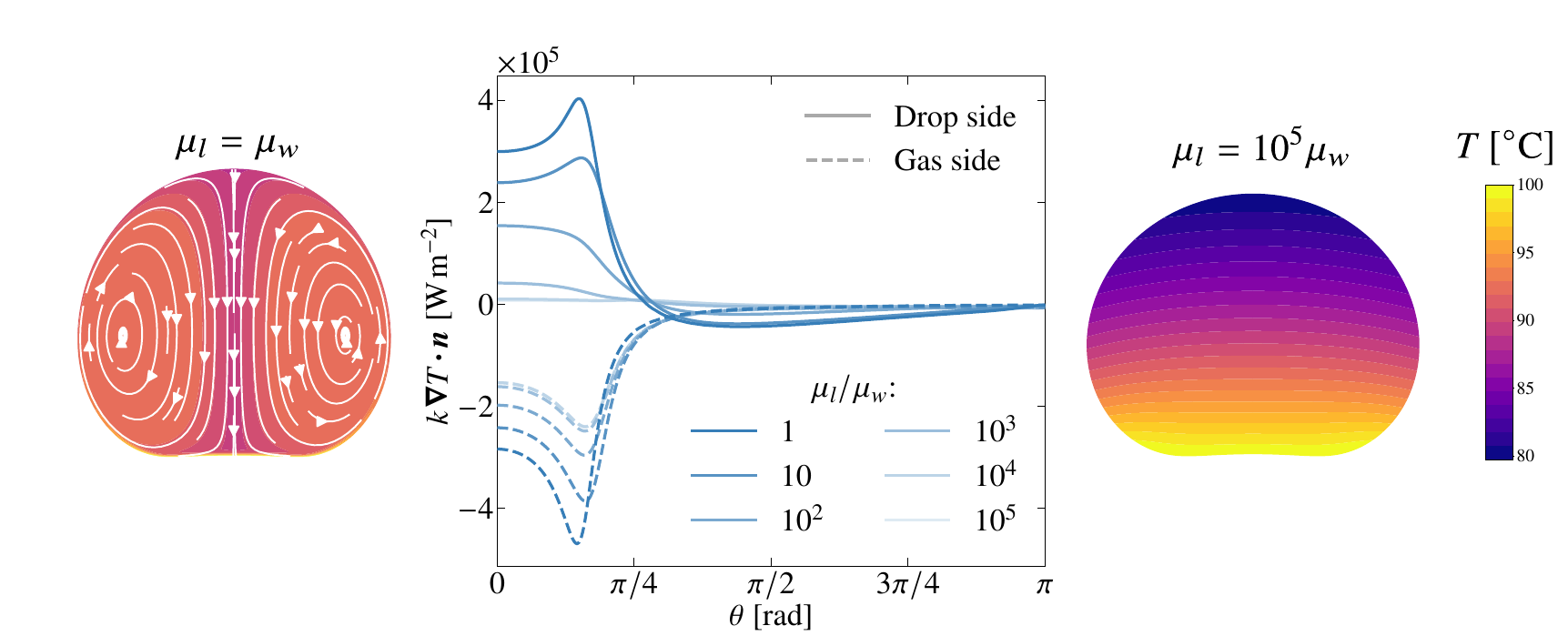}}
  \caption{Normal conductive heat flux from each side of the liquid-gas interface with increasing internal droplet viscosity for a droplet of size $R=1.59 \,\mathrm{mm}$ and substrate temperature $T_w=370\,^\circ\mathrm{C}$. 
  The sum of these two curves is proportional to the evaporation flux $j$. 
  Pictured outside the graphs are the axisymmetric temperature fields in each drop, in the two limiting case of internal droplet viscosity.}
\label{fig:tflux}
\end{figure}

Crucially, within the present model, the internal flows in the drop alter the thermal problem significantly by creating thin thermal boundary layers.
We can see this behaviour in figure \ref{fig:tflux} for a droplet of radius $R=1.59\,\mathrm{mm}$, where we increase the viscosity of the liquid in the mixed gas-vapour model. 
The sum of the two fluxes in the figure gives the jump condition \eqref{eqn:temp_jump}, which is proportional to the evaporation flux.
In the gaseous phase, when a thin vapour layer is present, the temperature gradient is approximately $\bnabla T\bcdot \boldsymbol{n} \sim (T_w-T_\mathrm{sat}(p_\mathrm{atm}))/h$ (this height changes with drop viscosity due to the coupled nature of the system). 
Inside the droplet, as the viscosity tends to that of water, the temperature gradients become non-negligible in comparison to those in the gaseous phase.
This is because cold liquid is advected down the axis and hot fluid is advected to the top of the droplet.
Consequently, due to the thermal gradients inside the droplet, the evaporation in the film layer decreases significantly and the contribution outside of this increases.

\section{Evaporation rate}\label{sec:evap}
The key to understanding the behaviour of Leidenfrost droplets is to investigate the evaporation kinetics. 
The simplest experimental measurement one can do for a Leidenfrost droplet is to measure the evaporation time. 
This is related to the total evaporation rate $J$, which can be obtained by integrating the evaporation flux around the surface of the droplet:

\begin{equation}
    J=\int_I j \,dS.
\end{equation}

In figure \ref{fig:m2p}, typical evaporation flux profiles around the surface of the droplet are shown.
The evaporative fluxes for models with neglected internal circulation have profiles similar to the pure vapour gaseous phase model (therefore we did not include them in the plot).
This is because the temperature gradients within the drop are negligible (or non-existent) in comparison to gradients in the gaseous phase.

\begin{figure}
  \centerline{\includegraphics[width=\textwidth]{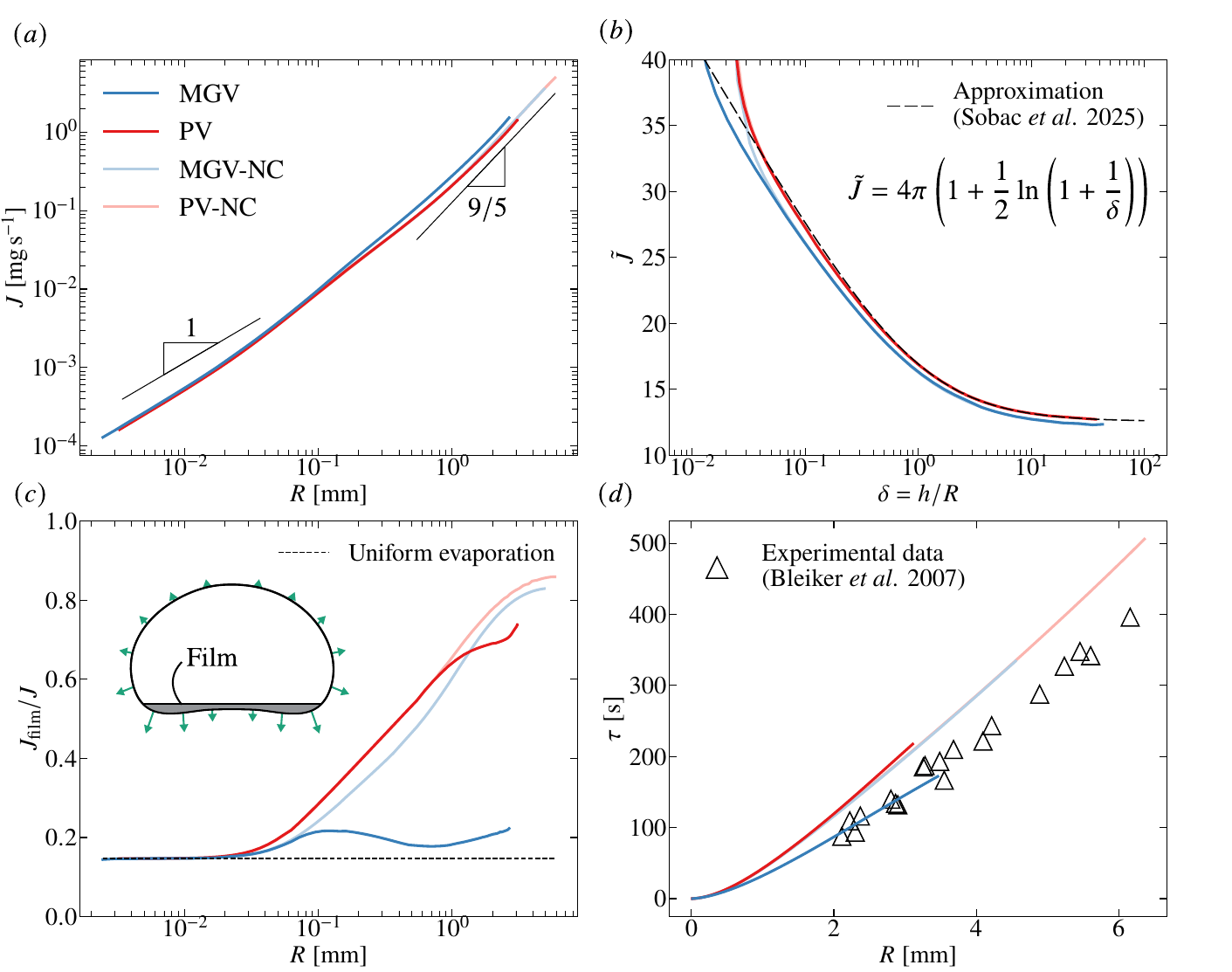}}
  \caption{
  Evaporation data for the models from figure \ref{fig:h}. 
  $(a)$ Log-log plot of the total evaporation rate against the volume of the drop. 
  $(b)$ Contribution to the total evaporation from the thin vapour film, the dashed line indicates the film contribution if the drop is spherical and evaporated uniformly. 
  The film layer is defined to begin from when the normal to the surface of the drop makes angle $\pi/4$ with the negative $z$ axis (for small spherical drops with uniform evaporation the contribution of film layer would be $\sin(\pi/8)^2$). 
  $(c)$ Non-dimensionalised evaporation versus non-dimensionalised drop height, compared to the data from theory of \citet{Sobac_Rednikov_Colinet_2025}. $(d)$ Integrated evaporation rates for a substrate temperature of $T_w=335\,^\circ \mathrm{C}$ compared to experimental data from \citet{BLEIKER2007835}.
  }
\label{fig:evap}
\end{figure}

In figure \ref{fig:evap}$(a)$, the total evaporation rate is plotted against the drop radius for each of the models in the previous section. 
Similarly to figure \ref{fig:h}, at the large drop limit, the evaporation rates of all models coincide apart from the model with both internal circulation and a mixed gas-vapour phase.
In contrast, in the lower limit, we see the models differing only due to humidity. 
Further, there are two distinct scaling laws in the upper and lower limits. 
The lower limit can be easily understood from isolated droplet evaporation since here the substrate is sufficiently far from the droplet that the only relevant length scale is the drop radius $R$:

\begin{equation}
    J \sim \frac{k_m\Delta T}{\mathcal{L}R}R^2 \sim R.
\end{equation}

\noindent This limit was studied in great detail by \citet{Sobac_Rednikov_Colinet_2025} who provided an approximation for the global evaporation rate in a pure vapour model, when non-dimensionalised in the following manner:

\begin{equation}\label{eqn:sobac_nondim}
    \delta \equiv h/R, \quad \tilde{J} \equiv J \left/\frac{k_m\Delta T}{\mathcal{L}}R\right.,
\end{equation}


\noindent where $\delta$ is the ratio of the droplet's height to its radius and $\tilde{J}$ is the dimensionless global evaporation rate.
The approximation is a simple form which respects the exact leading order asymptotic behaviours in the limits $\delta\rightarrow 0$ and $\delta\rightarrow \infty$.
In figure \ref{fig:evap}$(b)$, it can be seen that all curves roughly collapse onto the approximation derived in said paper (the equation in the figure).
The collapse is not perfect for the models considering a mixed gas-vapour gaseous phase due to the missing contributions of the temperature gradients within the droplet in the model of \citet{Sobac_Rednikov_Colinet_2025}. 
The non-dimensionalisations used for the numerical models are slightly nuanced due to their dependence on species mass fraction and temperature being considered in full.
As discussed in §\ref{sec:shapes}, the 1/3 rule is expected to provide the most accurate constant values for material properties of the models. 
Thus, the thermal conductivity of the gas $k_m$ is evaluated at:

\begin{equation}
    T = \frac{1}{3}T_w + \frac{2}{3}T_d, \quad w_v = \frac{2}{3}w_{v,d},
\end{equation}

\noindent where $T_d$ and $w_{v,d}$ are the surface averaged temperature and vapour mass fraction of the droplet, respectively. 
In equation \eqref{eqn:sobac_nondim}, the difference in temperature between the substrate and the drop is therefore given by $\Delta T = T_w - T_d$ and the latent heat is modified by the scale of the right hand side of the energy flux jump condition \eqref{eqn:temp_jump}.

In the limit of puddle shaped Leidenfrost droplets, as in the neglected circulation models in figure \ref{fig:shapes}, the scale for the global evaporation rate with droplet size may be obtained through lubrication theory.
This was first derived by \citet{POMEAU2012867} and also numerically obtained in \citet{SOBAC201585}. 
In this puddle configuration, we have a thin vapour layer and a droplet flattened by gravity with height twice the capillary length,

\begin{equation}
    \frac{4}{3}\pi R^3 = 2\ell_c\pi R_{\mathrm{max}}^2,
\end{equation}
\noindent implying $R_\mathrm{max} \sim R^{3/2}/\ell_c^{1/2}$. 
The horizontal length scale is simply the maximum radius, $L\sim R_\mathrm{max}$. 
If the height of the vapour layer $h$ is small ($h\ll L$), the reduced Reynolds number is small ($(h/L)\Rey_v=(h/L)\rho_vU_v h/\mu_v \ll 1$) and the Péclet number is small ($\Pen_v\ll1$), then $h$ is given to leading order by the quasi-stationary axisymmetric lubrication equation \citep{POMEAU2012867}:

\begin{equation}\label{eqn:axi_lub}
    \frac{1}{r}\frac{\p}{\p r}\left(-\frac{rh^3}{12\mu_v}\frac{\p p_v}{\p r}\right)=\frac{k_v\Delta T}{\rho_v\mathcal{L}h}, \quad p_v = p_0 + \rho_v g(h-z)- \gamma \left(\frac{\p^2 h}{\p r^2}+\frac{1}{r}\frac{\p h}{\p r}\right),
\end{equation}

\noindent where $p_v$ is the lubrication pressure in the vapour layer and internal drop circulation is neglected. 
The influence of gravity in the thin vapour is negligible since $\textit{Bo}_v = \rho_v g L^2/\gamma\ll1$. 
Therefore, the lubrication pressure is dominated by the capillary term in equation \eqref{eqn:axi_lub}, $p_v\sim \gamma h/L^2$.
From the lubrication equation, the pressure driving the flow must balance the evaporative flux, giving rise to the scale:

\begin{equation}
    p_v \sim \frac{k_v \Delta T \mu_v L^2}{\mathcal{L}h^4}.
\end{equation}

\noindent Balancing these pressures leads to a relation between the height $h$ in the thin layer and the initial drop radius $R$:

\begin{equation}
    h\sim \left(\frac{k_v \Delta T \mu_v R^6}{\gamma \mathcal{L}\ell_c^2}\right)^{1/5}\sim\left(\frac{\ell_{*}^3 R^6}{\ell_c^4}\right)^{1/5}.
\end{equation}

Finally, if we assume that the evaporation is dominated by the thin layer, which is verified in figure \ref{fig:evap}$(c)$, we recover the well-known $9/5$ scaling law as obtained in the numerical models shown in figure \ref{fig:evap}$(b)$:

\begin{equation}
    J\sim \frac{k_v\Delta T}{h}L^2\sim R^{9/5}.
\end{equation}

It is typically assumed that the evaporation is dominated by the vapour layer; however, in figure \ref{fig:evap}$(c)$, a quite different picture is obtained. 
In the small drop limit, all models approach uniform evaporation. 
Interestingly, in the large drop limit the mixed gas-vapour model with circulation  predicts that the vapour layer contribution does not completely dominate evaporation. 
This strongly contradicts the usual assumption that the film layer dominates evaporation of Leidenfrost droplets.
It is, therefore, peculiar that the scaling for global droplet evaporation of $J\sim R^{9/5}$ remains roughly respected despite key assumptions in this scenario being evidently incorrect.

As mentioned at the beginning of this section, the evaporation time is a quantity which is easily accessible from experiments. 
Despite our model being quasi-stationary, the drop lifetime may be obtained by integrating the global evaporation rate:

\begin{equation}
    \tau = -\int_V^0 \frac{\rho_l}{J(V')}\, dV'.
\end{equation}

Figure \ref{fig:evap}$(d)$ compares the predicted evaporation times from the models to experimental results from \citet{BLEIKER2007835} at a substrate temperature of $T_w = 335\,^\circ \mathrm{C}$.
The evaporation time of the mixed gas-vapour model with droplet circulation appears to best align with the experimental evaporation times, with other models over-predicting the time to evaporate quite significantly. 
Unfortunately, it is not possible to probe further into the large droplet regime with all of the models because of the unphysical nature of the droplet shapes due to azimuthal instability as discussed in §\ref{sec:stab}. 
Despite this limitation, it is clear that the global evaporation rate of the most complete model is systematically greater than that of the other simplified models due to the contribution from outside the thin layer.
As discussed in the previous section, this contribution arose from the significant thermal gradients within the droplet, particularly outside the thin layer.
Thus, it is paramount to consider the effect of both internal circulation and ambient humidity in order to accurately predict the evaporation time of Leidenfrost droplets.

It was shown in §\ref{sec:stab} that the mixed gas-vapour model with droplet circulation over-predicts the internal velocities of the droplets. 
In figure \ref{fig:h}, the $h_{\mathrm{neck}}$ data of \citet{burton_geometry_2012} sits above the mixed gas-vapour model with droplet circulation. 
The over-predicted internal velocities, enhancing the mechanism outlined in §\ref{sec:humcirc}, are thought to be a contributing factor to this discrepancy within the framework of the model.

\section{Further discussions of limitations of the model}\label{sec:extras}
The model presented in this paper allowed us to investigate and understand important mechanisms necessary to accurately model Leidenfrost droplets.
However, there are still several limitations of the current model which will be discussed in this section.
\subsection{Marangoni effects}\label{ssec:ma}
As stated in §\ref{sec:model}, a constant surface tension $\gamma$ is taken in the models thus far. 
In reality, however, the surface tension changes significantly with temperature.
This dependence can in general be well represented by a linear model:

\begin{equation}
    \gamma = \gamma_0 + \gamma_1 ( T- T_\mathrm{sat}(p_\mathrm{atm})).
\end{equation}

In the context of Leidenfrost droplets, thermo-capillary effects have been shown to have a large impact. 
Recently, \citet{MIALHE2023112366} presented a numerical method of an evaporating single component drop for which they showed a huge effect of thermal Marangoni driven flows on the droplet height which showed good agreement with the take-off results of \citet{celestini2012}. 
Furthermore, the model of \citet{aursand_thermocapillary_2018} predicts that thermal Marangoni forces facilitate film collapse of the vapour layer for certain film heights. 

\begin{figure}
  \centerline{\includegraphics[width=\textwidth]{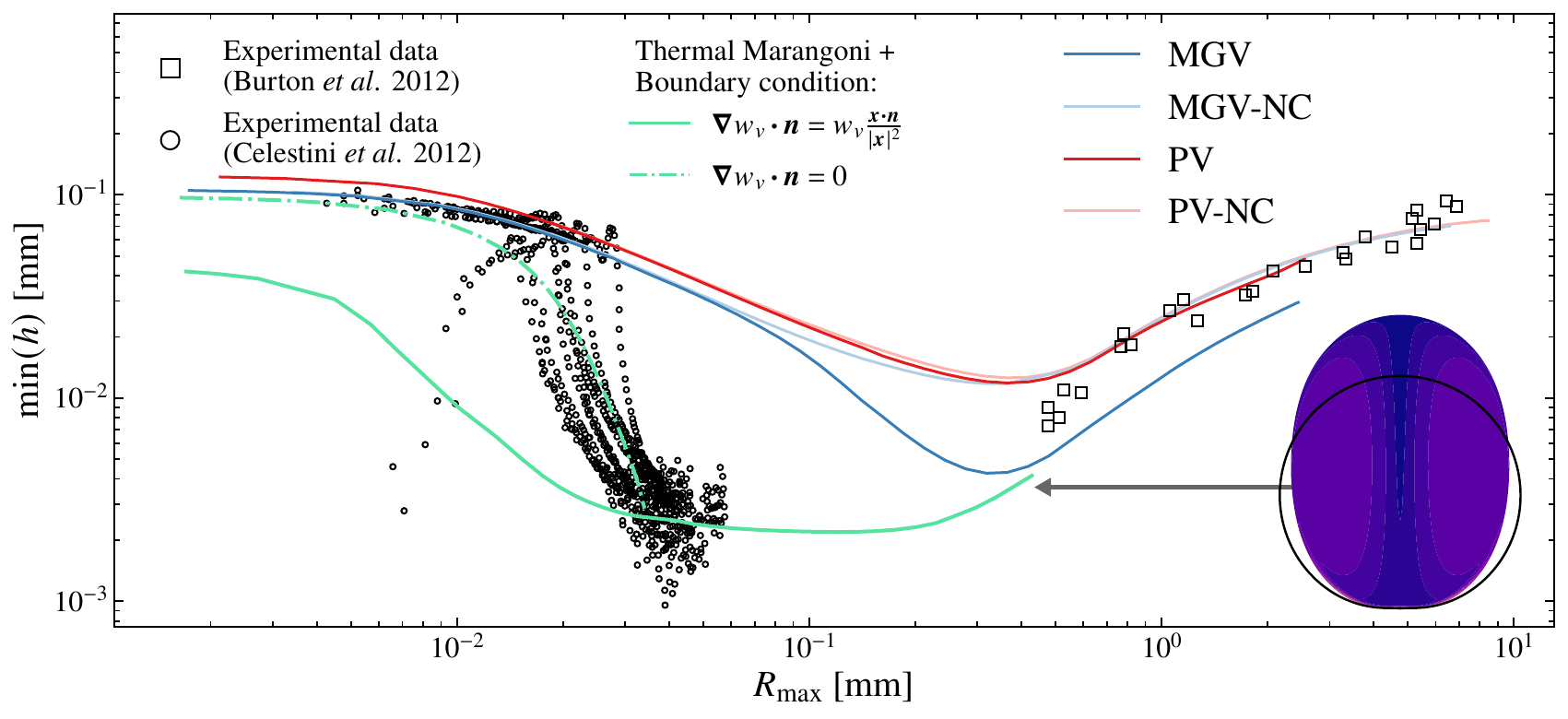}}
  \caption{Drop separation from the plate of the models in figure \ref{fig:h} with the new curve including the thermal Marangoni effect into the mixed gas-vapour gaseous model with circulation. 
  The inset shows the shape of the droplet with thermal Marangoni effect activated versus the shape of the mixed vapour model with the same drop volume at the rightmost data point on the thermal Marangoni curve.}
\label{fig:ma}
\end{figure}

In figure \ref{fig:ma}, we have replotted elements of figure \ref{fig:h} for comparison along with the full model with thermal Marangoni effects incorporated. 
The droplet separation from the plate decreases and finally the behaviour of the droplets closest to the substrate in the experiments of \citet{celestini2012} is captured.
This difference occurs since the drop is cooler at the top and therefore locally has higher surface tension. 
Hence, the flows due to thermal Marangoni forces are driven in the same direction as the shear from the vapour film. 
This exacerbates the mechanism outlined in the previous sections, further reducing the evaporation rates at the bottom of the drop due to large temperature gradients, causing the drop to sit closer to the plane. 
Simultaneously, this worsens the issue with prolate shapes by increasing the magnitude of flows in the droplet, as can be seen inset in figure \ref{fig:ma}. 
It is known that the flows driven by thermal Marangoni forces are often overestimated in numerical simulations in comparison to experiments. 
However, this does not account for the prolate drop shapes seen in the models of section §\ref{sec:shapes}. 

It can be seen that the drop separation from the plate of the experiments of \citet{celestini2012} falls between the curves with and without the thermal Marangoni effect.
In the context of this model, we understand the sudden take-off as the point where the flow inside the droplets becomes irrelevant and the drops sit at the same height as the models where internal flow is neglected.
A possible missing physical ingredient that could suppress circulation in the droplet is the presence of contaminants. 
For evaporating sessile droplets, a very small concentration of contaminants has been used to explain experimental observations by overwhelming the thermal Marangoni flow \citep{van_gaalen_competition_2022, rocha_evaporating_2025}. 
In the case of Leidenfrost drops, there have been many studies investigating the role of surfactants and the significant impact they can have on the final fate of a Leidenfrost drop which may explode before take-off such as in \citet{lyu_final_2019} and \citet{moreau_explosive_2019}. 
An avenue for future work would be to investigate whether the introduction of surfactants in this numerical model could be a missing component to explain experimental results which fall between our different models in figure \ref{fig:ma}.

It was mentioned in §\ref{sec:method}, that particular care should be taken with the boundary condition, since the natural condition of $\bnabla w_v\bcdot \boldsymbol{n} = 0$ causes the domain to incorrectly fill with a significant amount of vapour (the value of which is even domain dependent). 
This reduces the cooling effect and therefore temperature gradients in the droplet.
Hence, the vertical evaporative force on the drop is increased and the height approaches that of the models in §\ref{sec:shapes}. 
In figure \ref{fig:ma}, the mint dash-dotted line coincidentally appears to show better agreement with the experimental data of \citet{celestini2012}.
However, it should be made clear that any mechanism through which temperature gradients within the droplet are reduced would produce such an effect.

\subsection{Axisymmetry}\label{ssec:axisym}

In §\ref{sec:stab}, we showed that the axisymmetric models with internal circulation become azimuthally unstable above a critical radius of $R\approx 0.2\,\mathrm{mm}$.
The reason for prolate drop shapes within the axisymmetric model may therefore be due to a combination of two effects: azimuthally averaged flows having lower velocities than what is predicted in the axisymmetric simulation and the flow having the ability to move in the azimuthal direction thus not drastically augmenting the geometry of the free surface.

The effects of symmetry-breaking flows may be investigated using 2D analogues (no axisymmetry) of the present model, such as in \citet{brandao_spontaneous_2020}, as well as in \citet{gauthier_self-propulsion_2019} to better understand the inverse Leidenfrost effect.
However, in order to prove that the onset of the instabilities of §\ref{sec:stab} materialises into this large effect on the droplet shape, 3D simulations must be used.
For the model presented in this paper, full 3D direct numerical simulations are computationally challenging at present.
However, we will support the hypothesis that axisymmetry plays the main role in this discrepancy using the coupled Navier-Stokes to lubrication stress model presented in the work of \citet{Chakraborty_Chubynsky_Sprittles_2022}.

On the vapour layer region of the drop, we solve the following more general lubrication equation than that of equation \eqref{eqn:axi_lub} in §\ref{sec:evap}:

\begin{equation}
    \frac{\p h}{\p t} + \bnabla\bcdot \left(-\frac{h^3}{12\mu_v}\bnabla p_v+\frac{h}{2}\boldsymbol{u}_{l}\right) = \frac{k_v\Delta T}{\rho_v\mathcal{L}h},
\end{equation}

\noindent where $h$ is the height of the layer and $p_v$ is the lubrication pressure in the vapour layer.
The stresses from the thin layer are then coupled to the Navier-Stokes equation in the droplet through the dynamic boundary condition:

\begin{equation}
    \boldsymbol{\sigma}_l\bcdot\boldsymbol{n} = -\gamma\left(\bnabla_S\bcdot\boldsymbol{n}\right)\boldsymbol{n} - \left(p_v\boldsymbol{n} +\frac{h}{2}\bnabla p_v +\frac{\mu_v}{h}\boldsymbol{u}_l\right),
\end{equation}

\noindent with the patching point for the lubrication equations chosen where the outward-facing normal makes an angle of $\pi/4$ with the $z$ axis.
In classical lubrication theory, the gradients and velocities in the previous two equations should be taken in the horizontal direction.
However, we take them in the direction of the surface for numerical simplicity. 
This introduces an error on the order of $\mathcal{O}\left(|\boldsymbol{n}-\boldsymbol{e}_z|\right)$, which is insignificant for the following qualitative results.

\begin{figure}
  \centerline{\includegraphics[width=\textwidth]{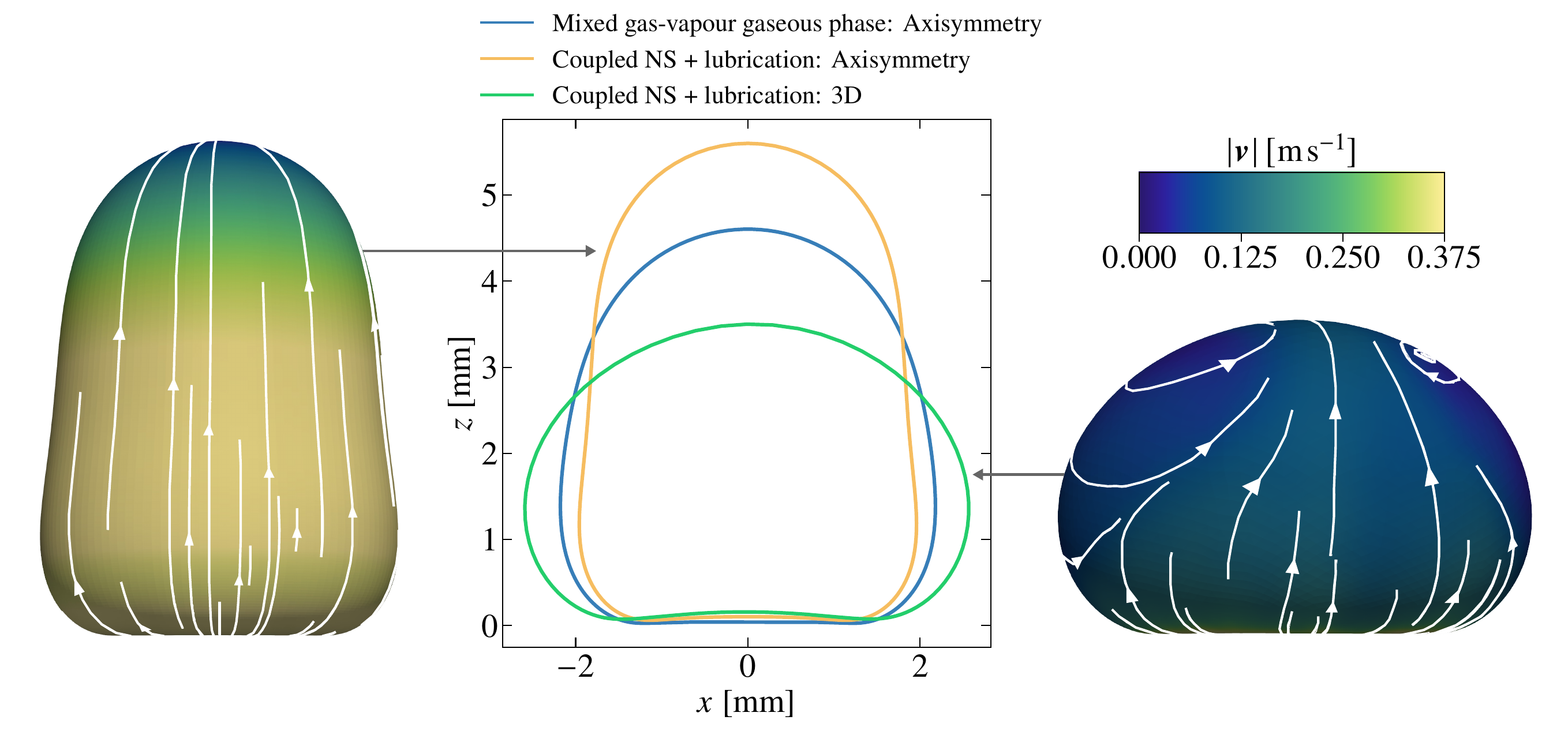}}
  \caption{Surface contours of a $R=2.3\,\mathrm{mm}$ Leidenfrost droplet above a plate at $T_w=370\,\mathrm{^\circ C}$ in the plane $y=0$. 
  The models plotted are the mixed gas-vapour gaseous phase model with drop circulation from this work against the coupled Navier-Stokes lubrication models in axisymmetry and 3D (with a plane of symmetry in $y=0$). 
  Each model was run until a stable droplet shape is reached.
  Adjacent to the graph are projections of 3D simulations of the lubrication based models with surface streamlines and colour bar displaying the magnitude of the velocity.}
\label{fig:axisym}
\end{figure}

The result of running the above numerical model for a Leidenfrost drop with $R\approx \ell_c$ in axisymmetry and on a 3D hemisphere is shown in figure \ref{fig:axisym}. 
In the 3D case, a half drop is simulated with symmetry in the plane $y=0$ to reduce the computational demand while still permitting the formation of all azimuthal modes.
In axisymmetry, the lubrication based model produces even more prolate shapes than the models presented in this paper.
Yet, loosening the constraint of axisymmetry in this model dramatically affects the drop contour, returning to the expected puddle-like forms.
This strongly supports our hypothesis that axisymmetry may not be assumed in numerical simulations of large Leidenfrost droplets.

\section{Summary and conclusions}\label{sec:conc}
In this work, we used direct numerical simulations to model Leidenfrost droplets hovering above an isothermal plate. 
The simulations were used to analyse droplets over four decades of drop size spanning from puddles to spherical droplet take-off. 

We analysed the mixed gas-vapour gaseous phase model against simplified models through varying the far-field ambient humidity of the gaseous phase and the viscosity inside the droplet. 
Specifically, we showed that, within the context of the model, consideration of both internal drop circulation and a mixed gas-vapour gaseous phase makes a significant difference to the nature of the evaporation process of droplets with $R\gg\ell_*$. 
The frequently made assumption that evaporation is dominated by the underside of the droplet appears not to be valid in this model. 
Instead, the presence of large temperature gradients in the droplet reduces the evaporation at the base of the drop and increases the contribution from the outer drop surface. 
Overall, this has the effect of increasing the global evaporation rate while allowing the drop to sit closer to the plane, agreeing with elements of experimental results that are not explained by lubrication models. 
Without this consideration, numerical models massively over-predict the lifetime of Leidenfrost droplets compared to those measured in experiments.
This highlights the importance of the coupled role of both ambient humidity and internal droplet circulation.

Throughout the analysis of the numerical models, we recovered predicted scaling laws for the drop height $h\sim R^{-1/2}$ from \citet{POMEAU2012867} and \citet{celestini2012} near the take-off regime, as well as scaling laws for the global evaporation rate in the large and small drop limits of $J\sim R^{9/5}$ for $R\gtrsim\ell_c$ and $J\sim R$ for $R\lesssim \ell_*$ from \citet{POMEAU2012867}.

The consideration of simplified models with a pure vapour gaseous phase and the neglect of internal drop motion allowed us to connect theoretical works of \citet{sobac_leidenfrost_2014} and \citet{Sobac_Rednikov_Colinet_2025} in one complete model, spanning the entire lifetime of stable Leidenfrost droplets.

Although much was learnt about the importance of ambient humidity and internal drop circulation, there are several limitations to the presented axisymmetric model.
In the case of large Leidenfrost droplets, we observed a surprisingly worse agreement of the shape of the `more complete' models with experimental measurements.
This is attributed to the onset of an azimuthal instability for droplets with $R\gtrsim 0.2\,\mathrm{mm}$.
We showed, using the coupled Navier-Stokes to lubrication stress model (which also produces prolate drop shapes), that axisymmetry is the critical invalid assumption made for the discrepancy between the numerical models and experiments.
As shown in this paper, consideration of internal drop circulation is necessary to accurately predict drop lifetimes. 
Since the internal drop circulation causes azimuthal instability, this means that axisymmetric models of large Leidenfrost drops ($R\gtrsim\ell_c$) with drop circulation are questionable.

Within the regime where the axisymmetric model is valid, improvements would involve the inclusion of thermal Marangoni forces as well as contamination via surfactants.
Extra care must be taken here as the take-off is not quasi-stationary (due to the fast change of geometry) and it is expected that the thermal Marangoni driven flows may trigger instability earlier.
To aid in elucidating which physical effects are at play here, we encourage experimental analysis for droplet geometry to be taken, in a regime in between the results of \citet{celestini2012} and \citet{burton_geometry_2012}, namely in the regime where the droplet is closest to the surface and $\ell_*\lesssim R\lesssim \ell_i$.

\begin{bmhead}[Acknowledgements.]
This project has received funding from the European Union’s Horizon Europe research and innovation programme under the Marie Sk\l odowska-Curie Actions (MSCA) grant agreement No. 101169365. The views and opinions expressed are however those of the author(s) only and do not necessarily reflect those of the European Union or the European Research Executive Agency (REA). Neither the European Union nor the granting authority can be held responsible for them.
\end{bmhead}

\begin{bmhead}[Declaration of interests.]
The authors report no conflict of interest.
\end{bmhead}

\begin{appen}

\section{}\label{appA}

\subsection{Computational details}\label{appA:comp}
Details of the numerical implementation in \textsc{pyoomph} \citep{diddens_bifurcation_2024} are provided in the following section.

In order to obtain the quasi-stationary solutions for Leidenfrost droplets, a transient simulation of an initially spherical droplet is run until a stationary state is reached by solving the system of equations in §\ref{ssec:gov} (MGV). 
The mesh is made using second order unstructured triangular elements.
The Taylor-Hood element pair is used for velocity and pressure and for all other scalars second order Lagrange (C2) elements are used.
The liquid-gaseous phase interface may move and is solved for with an arbitrary Lagrangian-Eulerian method.
The nodes of the mesh are then evolved by solving the linear static elasticity equation.
The time stepping used is the second-order backward differentiation formula and time steps are chosen automatically by ensuring temporal error is lower than a given threshold. 
We employ a numerical trick of linearly increasing gravity to gently deposit the Leidenfrost droplet on the surface, which aids significantly in speed of convergence.
Since solutions are quasi-stationary, the volume of the drop is kept constant by enforcement using a Lagrange multiplier for the pressure.

The final stationary state of the triangular mesh is used as template to create a mesh using second order quad Lagrange elements, where a-posteriori spatial mesh adaptivity based on Zienkiewicz-Zhu (Z2) error estimation is used to refine elements in areas of large gradients in velocity, temperature and composition.
The simplified models are obtained using arc length continuation in the required parameter: $\mu_l$ to neglect internal drop circulation and $w_\mathrm{amb}$ for the pure vapour gaseous phase.
Similarly, arc length continuation is utilised with the volume of the drop to efficiently acquire stationary solutions for different drop sizes.
Whenever the quality of mesh elements falls below a threshold, the domain is remeshed and variables are interpolated onto the new mesh.

For the take-off Leidenfrost drops, where transient simulations must be utilised the method is the same as above, with the exception that there is no Lagrange multiplier constraint on the volume. 
Therefore, the drop may evaporate and reduce in size.
All solutions on different domains are then stitched together to create the final curves in this manuscript.

The domain size was chosen to be sufficiently large that increasing the domain size further has negligible effect on the near drop quantities.
Similarly, the mesh size was chosen to be small enough that upon further refinement there is no graphical difference between the curves.

In order to conduct the azimuthal stability analysis of §\ref{sec:stab}, all equations must be expanded to linear order in the perturbation parameter $\varepsilon$. 
\textsc{pyoomph} does this computation symbolically. 
To assist the symbolic computation of expressions in the \textsc{GiNaC} core, the material properties in the gaseous domain are taken at the film temperature $(T_w+T_\mathrm{sat}(p_\mathrm{atm}))/2$ and the vapour saturation pressure \eqref{eqn:vap_sat_p} is Taylor expanded about the saturation temperature to fourth order (sufficient to ensure accuracy over the required temperature range).

\subsection{Thermophysical properties}\label{appA:props}

\begin{table}
  \begin{center}
\def~{\hphantom{0}}
  \begin{tabular}{cccccc}
      Properties & Units & Liquid & Vapour & Gas & Interface \\[3pt]
      $\rho$ & $\mathrm{kg \,m^{-3}}$ & 958.35 & - & - & - \\
      $\mu$ & $\mathrm{mPa \,s^{-1}}$ & 0.28559 & \eqref{eqn:mu_v} & \eqref{eqn:mu_g} & - \\
      $k$ & $\mathrm{W \,m^{-1}\,K^{-1}}$ & 0.6791 & \eqref{eqn:k_v} & \eqref{eqn:k_g} & - \\
      $c_p$ & $\mathrm{J \,kg^{-1}\,K^{-1}}$ & 4217 & 2077 & 1011 & - \\
      $D_{vg}$ & $\mathrm{m^2 \,s^{-1}}$ & - & \eqref{eqn:D_vg}&\eqref{eqn:D_vg}& -\\
      $M_i$ & $\mathrm{g\, mol^{-1}}$ & -& 18.02 & 28.97 &  -\\
      $\gamma_0$ & $\mathrm{mN\, m^{-1}}$ &- &- &- & 58.9 \\
      $\gamma_1$ & $\mathrm{mN\, m^{-1}}\,\mathrm{K}^{-1}$ &- &- &-& 0.185 \\
      $\mathcal{L}$ & $\mathrm{kJ\, kg^{-1}}$ & -& -&-& 2257 \\
      $T_\mathrm{sat}$ & K & -& -& -& 373.15 \\
      $p_\mathrm{atm}$ & atm & -& -& 1 &- \\
      $R_u$ & $\mathrm{J\, mol^{-1}\, K^{-1}}$ &- & -& 8.314 & -\\
  \end{tabular}
  \caption{Thermophysical properties of liquid water, vapour and dry air at atmospheric pressure used in the numerical model.}
  \label{tab:mat_props}
  \end{center}
\end{table}

The thermophysical properties are listed in table \ref{tab:mat_props}.
The temperature dependence of the material properties in the gaseous phase is plotted in figure \ref{fig:mat_props}. The linear lines of best fit of each of the material properties are given below:

\begin{figure}
  \centerline{\includegraphics[width=\textwidth]{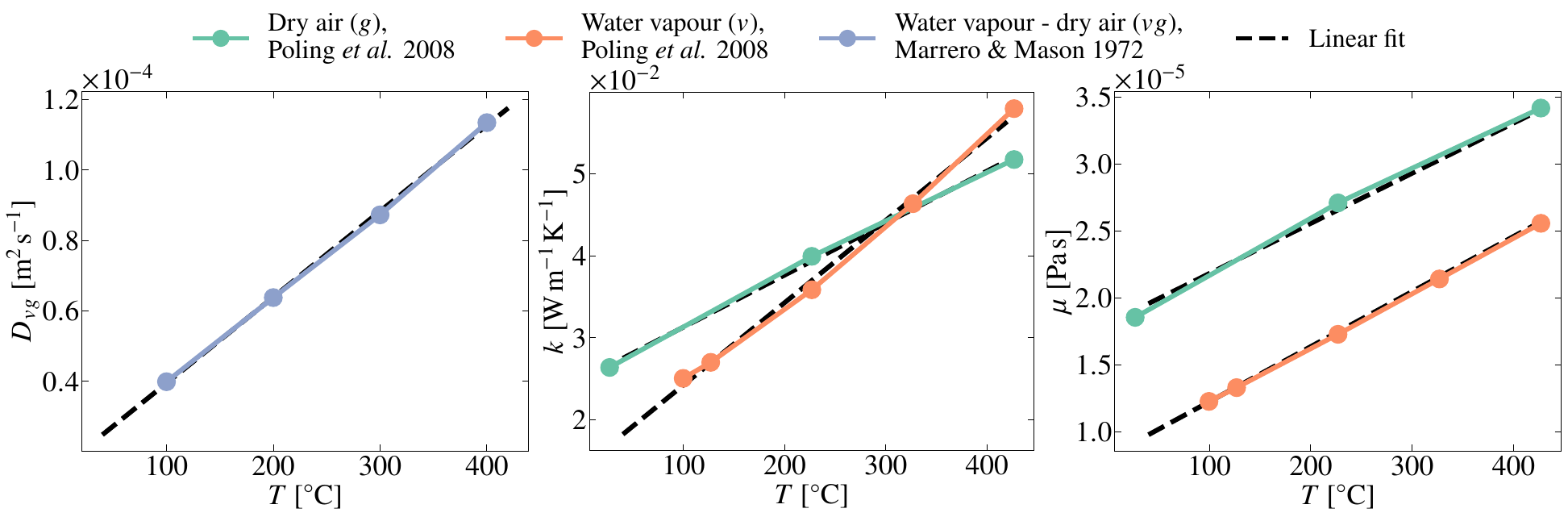}}
  \caption{Variation of thermophysical properties with temperature in the gaseous phase of each of dry air and water vapour. 
  The experimental data for thermal conductivity and dynamic viscosity are taken from \citet{perrys2008_physchemdata} and the data for the binary diffusion coefficient comes from \citet{Marrero}.
  }
\label{fig:mat_props}
\end{figure}

\begin{equation}\label{eqn:mu_v}
    \mu_v = (4.12319477 \times 10^{-8} \,T/\mathrm{K} - 3.16122622\times10^{-6})\, \mathrm{Pa \,s^{-1}}
\end{equation}
\begin{equation}\label{eqn:mu_g}
    \mu_g = (3.69506631 \times 10^{-8} \,T/\mathrm{K} + 8.16777909\times10^{-6})\, \mathrm{Pa \,s^{-1}}
\end{equation}
\begin{equation}\label{eqn:k_v}
    k_v = (1.0006 \times 10^{-4} \,T/\mathrm{K} -0.01353965)\, \mathrm{W \,m^{-1}\,K^{-1}}
\end{equation}
\begin{equation}\label{eqn:k_g}
    k_g = (6.23484729 \times 10^{-5} \,T/\mathrm{K} -8.53176446\times10^{-3})\, \mathrm{W \,m^{-1}\,K^{-1}}
\end{equation}
\begin{equation}\label{eqn:D_vg}
    D_{vg} = (2.4430000 \times 10^{-7} \,T/\mathrm{K} - 5.1680545\times10^{-5})\, \mathrm{m^2 \,s^{-1}}
\end{equation}

\noindent where the maximum relative error between the linear fit and given data does not exceed 3\%.

\subsection{Validity of quasi-stationary assumption}\label{ssec:validityquasi}
In order to neglect the time derivatives and assume a quasi-stationary model in §\ref{ssec:quasi}, we must ensure that other relevant time scales in the problem are negligible in comparison to the evaporative time scale of the drop. 
The relevant time scales are the hydrodynamic, the thermal, and the dynamic timescale. However, the dynamic timescale is only relevant for take-off drops where the geometry changes on a comparable timescale to the evaporation time.
This regime was discussed in §\ref{sec:shapes}.

For liquid water at around $100\,^\circ \mathrm{C}$, as well as for the gaseous mixture over a reasonable temperature range outside those used in this study, the Prandtl number is $\Pr = \mu c_p/k\sim \mathcal{O}(1)$ in each respective domain. 
Thus it suffices to only consider one of the thermal or hydrodynamic - we will consider the hydrodynamic scales.

There are two distinct geometries to consider to establish the quasi-stationary assumption across the lifetime of the Leidenfrost droplet. 
The first is for scenarios where a thin vapour layer is present and the drop is flattened at the bottom $R\gtrsim \ell_i$ and the second is when the droplet is much smaller than this, becomes spherical and evaporates an appreciable distance from the substrate. 

The timescale for evaporation is given by  $\tau_\mathrm{ev}=\rho_l R^3/J$, where $J\sim jL^2$ is the integrated evaporation flux around the droplet and $L$ is the length scale of the area over which evaporation dominates. 

\subsubsection{Small Leidenfrost drops}
In the case of small spherical drops, approximately given by $R\sim 50\,\upmu\mathrm{m}$, the only relevant length scale is the radius $R$. 
Thus, the evaporative flux is $j\sim k_m \Delta T /\mathcal{L} R$ with $L\sim R$, since the evaporation is roughly uniform. 
This case is covered in detail in \citet{Sobac_Rednikov_Colinet_2025}. The evaporation time, therefore, roughly scales as:

\begin{equation}
    \tau_\mathrm{ev}\sim\frac{\rho_l \mathcal{L} R^2}{k_m \Delta T}\sim 10^{-1}\,\mathrm{s}.
\end{equation}

\noindent Due to the small droplet size, the Reynolds numbers in each phase are small and thus the viscous relaxation timescale is of importance.

\begin{equation}
    \tau_\mathrm{vis}^m \sim \frac{\rho_mR^2}{\mu_m}\sim 10^{-5}\,\mathrm{s}, \quad \tau_\mathrm{vis}^l \sim \frac{\rho_lR^2}{\mu_l}\sim 10^{-2}\,\mathrm{s}.
\end{equation}

\noindent Hence, we have that $\tau_\mathrm{vis}^m,\tau_\mathrm{vis}^l\ll \tau_\mathrm{ev}$ for this size of small Leidenfrost drops.

\subsubsection{Large Leidenfrost drops}
In the case of larger drops, approximately given by $R\sim 5\,\mathrm{mm}, \, h\sim100 \,\upmu\mathrm{m}$, the vapour layer is well-known to be quasi-stationary through arguments in lubrication theory. 
This remains valid so long as the reduced Reynolds number of the vapour layer is small $(h/R) Re_v\ll 1$, where $Re_v\sim \rho_v U_v h/\mu_v$. 
The vapour flow underneath the droplet is driven by evaporation. 
Following the arguments in the lubrication theory of \citet{POMEAU2012867}, one obtains the scales:

\begin{equation}\label{eqn:lub_vap_scaling}
    U_v \sim \frac{R}{h}\frac{j}{\rho_v}, \quad j\sim\frac{k_v \Delta T }{\mathcal{L} h}, \quad L\sim \left(\frac{R^3}{\ell_c}\right)^{1/2}.
\end{equation}

\noindent Thus, the evaporative time scale is:

\begin{equation}
    \tau_\mathrm{ev}\sim\frac{\rho_l \mathcal{L} h\ell_c}{k_v \Delta T}\sim 100\,\mathrm{s}.
\end{equation}

\noindent We now look at the flow inside the droplet, which is driven by shear with the vapour layer.
From continuity of shear stress at the interface, we have $\mu_l U_l/R\sim \mu_v U_v/h$.
Thence, using equation \eqref{eqn:lub_vap_scaling}, the Reynolds number inside the drop is for large drops of order:

\begin{equation}
    Re_l \sim \frac{\rho_lU_l R}{\mu_l}\sim\frac{\mu_v\rho_l j}{\mu_l^2\rho_v}\frac{R^3}{ h^2} \sim \mathcal{O}(10^3).
\end{equation}

\noindent The outer component of the gas instead has a characteristic velocity governed by the evaporative flux only, since the flow is not confined to a thin geometry $U_m\sim j/\rho_m$. 
Thus, the Reynolds number is:

\begin{equation}
    Re_m\sim \frac{\rho_m U_m R}{\mu_m}\sim \mathcal{O}(10).
\end{equation}

\noindent Therefore, in both scenarios, the relevant hydrodynamic time scale is the advective timescale:

\begin{equation}
    \tau_\mathrm{adv}^l\sim R/U_l\sim 10^{-3}\,\mathrm{s}, \quad \tau_\mathrm{adv}^m\sim \frac{R}{U_m}\sim 10^{-2}\mathrm{s}.
\end{equation} 

\noindent Thus, in the case of large drops, we have $\tau_\mathrm{adv}^m,\tau_\mathrm{adv}^l\ll\tau_\mathrm{ev}$. 
Indeed, in both regimes all important timescales in the problem are much less than the respective evaporative time scales.
It is assumed that between these limits the system remains quasi-stationary, thus justifying the assumption made in §\ref{ssec:quasi}. 
This assumption is also self verified by the results of the numerical simulation.

\end{appen}\clearpage

\bibliographystyle{jfm}
\bibliography{deWildt_2025}

\end{document}